\documentclass[sigplan,10pt]{acmart}

\renewcommand\footnotetextcopyrightpermission[1]{}
\usepackage{subcaption}

\usepackage{booktabs}
\usepackage{multirow}
\usepackage{etoolbox}
\usepackage{xcolor}
\usepackage{caption}
 \usepackage{xspace}
\usepackage{amsfonts}
\usepackage{mathtools}
\usepackage{tabularx}
\usepackage{enumitem}
\usepackage{float}
\usepackage[font=bf,labelfont=small]{caption}
\usepackage[binary-units=true]{siunitx}
\usepackage{pgfplots}
\usepackage{outlines}
\usepackage{comment}
\usepackage{balance}

\usepackage{algorithm}
\usepackage{algpseudocode}

\newcounter{packednmbr}

\newenvironment{packeditemize}{\begin{list}{$\bullet$}{\setlength{\itemsep}{0.5pt}\addtolength{\labelwidth}{-4pt}\setlength{\leftmargin}{\labelwidth}\setlength{\listparindent}{\parindent}\setlength{\parsep}{1pt}\setlength{\topsep}{2.5pt}}}{\end{list}}

\newbool{IsPrintComment}
\booltrue{IsPrintComment}   

\newcommand{\name}{PrefillOnly\xspace}

\newcommand{\todo}[1]
{
	\ifbool{IsPrintComment}
	{
		{\color{blue} #1}
	}
	\  
}

\newcommand{\kt}[1]
{
	\ifbool{IsPrintComment}
	{
		{\color{teal}(KD) #1}
	}
	\  
}

\newcommand{\zc}[1]
{
	\ifbool{IsPrintComment}
	{
		{\color{olive}(Chen) #1}
	}
	\  
}
\newcommand{\lily}[1]
{
	\ifbool{IsPrintComment}
	{
		{\color{blue}(Lily) #1}
	}
	\  
}

\newcommand{\yq}[1]
{
	\ifbool{IsPrintComment}
	{
		{\color{orange}(Yifan) #1}
	}
	\  
}

\newcommand{\hejian}[1]
{
	\ifbool{IsPrintComment}
	{
		{\color{green}(Hejian) #1}
	}
	\  
}

\newcommand{\ion}[1]
{
	\ifbool{IsPrintComment}
	{
		{\color{brown}(Ion) #1}
	}
	\  
}

\newcommand{\TODO}[1]
{
	\ifbool{IsPrintComment}
	{
		{\color{red}todo: #1}
	}
	\  
}

\definecolor{hhcolor}{HTML}{EE22EE}
\newcommand{\hh}[1]
{
	\ifbool{IsPrintComment}
	{
		{\color{hhcolor}(Yihua) #1}
	}
	\  
}

\newcommand{\squishlist} 
{
    \begin{list}{$\bullet$}
    {
        \setlength{\itemsep}{0pt}      \setlength{\parsep}{3pt}
        \setlength{\topsep}{3pt}       \setlength{\partopsep}{0pt}
        \setlength{\leftmargin}{1.5em} \setlength{\labelwidth}{1em}
        \setlength{\labelsep}{0.5em}
    }
}

\newcommand{\squishend}
{
    \end{list}
}

\newcommand{\eg}{{\it e.g.,}\xspace}
\newcommand{\ie}{{\it i.e.,}\xspace}

\newcommand{\mypara}[1]{\vspace{0.04cm}\noindent{\bf {#1}:}~}

\newcommand{\tightcaption}[1]{
\caption{{\normalfont{\textit{{#1}}}}}
}

\newcommand{\tightsection}[1]{
\section{#1}
}

\begin{document}

\pagestyle{plain}


\title{\name: An Inference Engine for Prefill-only Workloads in Large Language Model Applications}



\author{Kuntai Du\textsuperscript{‖}, Bowen Wang\textsuperscript{†}, Chen Zhang\textsuperscript{†}, Yiming Cheng\textsuperscript{‖}, Qing Lan\textsuperscript{‡}, 
Hejian Sang\textsuperscript{‡}, 
Yihua Cheng\textsuperscript{‖},
Jiayi Yao\textsuperscript{‖},
Xiaoxuan Liu\textsuperscript{¶},
Yifan Qiao\textsuperscript{¶},
Ion Stoica\textsuperscript{¶},
Junchen Jiang\textsuperscript{‖} \\
\vspace{0.2cm}
\textsuperscript{‖}University of Chicago \quad 
\textsuperscript{†}Tsinghua University \quad 
\textsuperscript{‡}LinkedIn \quad 
\textsuperscript{¶}UC Berkeley
}






\begin{abstract}

Besides typical generative applications, like ChatGPT, GitHub Copilot, and Cursor, we observe an emerging trend that LLMs are increasingly used in traditional discriminative tasks, such as recommendation, credit verification, and data labeling. 
The key characteristic of these emerging use cases is that the LLM generates only a {\em single} output token, rather than an arbitrarily long sequence of tokens. 
We call this {\em prefill-only} workload.
However, since existing LLM engines assume arbitrary output lengths, they fail to leverage the unique properties of prefill-only workloads. 
In this paper, we present \name, the first LLM inference engine that improves the inference throughput and latency by fully embracing the properties of prefill-only workloads. 
First, since it generates only one token, \name only needs to store the KV cache of only the last computed layer, rather than of all layers. This drastically reduces the GPU memory footprint of LLM inference and allows handling long inputs without using solutions that reduces throughput, such as cross-GPU KV cache parallelization.
Second, because the output length is fixed, rather than arbitrary, \name can precisely determine the job completion time (JCT) of each prefill-only request before it starts. 
This enables efficient JCT-aware scheduling policies such as shortest remaining job first. 
\name can process upto $4\times$ larger queries per second without inflating average and P99 latency.


\end{abstract}

\settopmatter{printfolios=true}
\maketitle



\section{Introduction}

Nowadays, large language models (LLMs) are widely used in generative tasks, such as chatting (\eg ChatGPT~\cite{chatgpt}, Character.AI~\cite{characterai}), code generation (\eg GitHub Copilot~\cite{copilot}, Cursor~\cite{cursor}), and summarization (\eg Perplexity~\cite{perplexityai}), which generate new content (of variable lengths) for people to read and use.
However, researchers and industry practitioners also recently started to use LLMs to replace traditional deep learning models in {\em \bf discriminative} tasks, such as recommendation~\cite{firooz2025360brew,wang2023enhancing,wu2024survey}, credit verification~\cite{CatBoost,CALM,Vicuna}, and data labeling~\cite{Annollm,llmaaa,Doris}.
The reason is two-fold: 
\begin{packeditemize}
    \item \textit{Streamlining development}: 
    Traditional deep learning solutions often require multiple iterations of data pre-processing~\cite{PreProAI}, feature engineering~\cite{feature}, and model tuning~\cite{modeltune}, which involves cross-team collaboration and can be time-consuming. 
    In contrast, LLMs can directly process raw data and are general enough to obviate fine-tuning~\cite{firooz2025360brew,gpt3,limitedGenAI}, allowing developers to interactively debug and improve quality by just changing the prompts. 
    \item \textit{High decision quality}: By carefully choosing the LLMs and engineering the prompt of the LLM-based pipeline, LLMs can achieve comparable or even higher decision quality compared to production models~\cite{firooz2025360brew}.
\end{packeditemize}

\begin{figure}
    \centering
    \begin{minipage}{1\columnwidth}
    \centering
    \includegraphics[width=\linewidth]{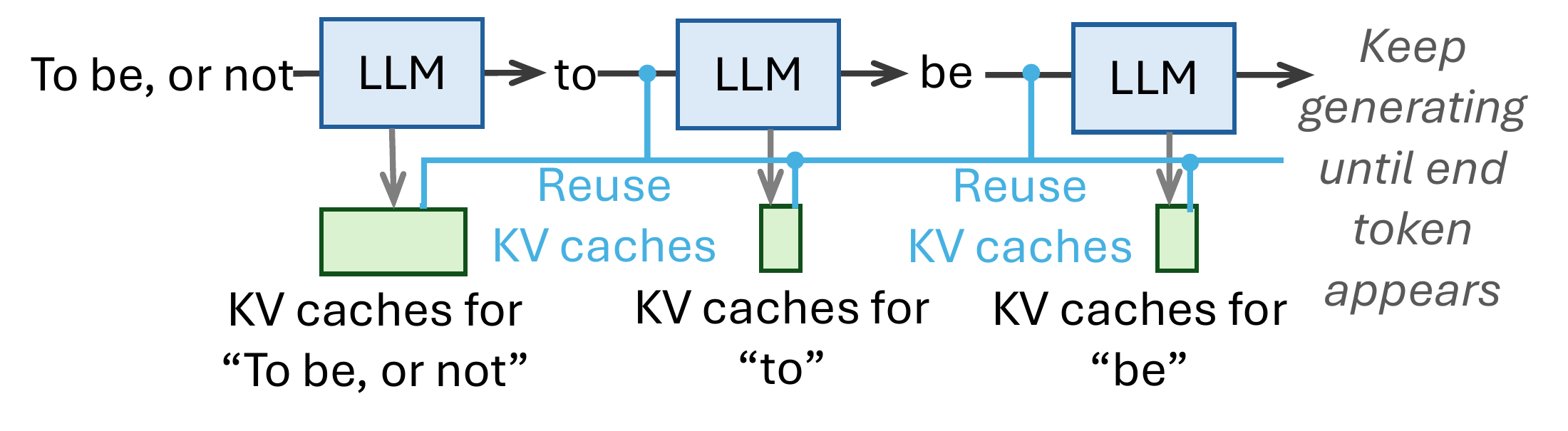}
    \subcaption{Traditional generative LLM inference}
    \end{minipage}

    \begin{minipage}{0.8\columnwidth}
    \centering
    \includegraphics[width=\linewidth]{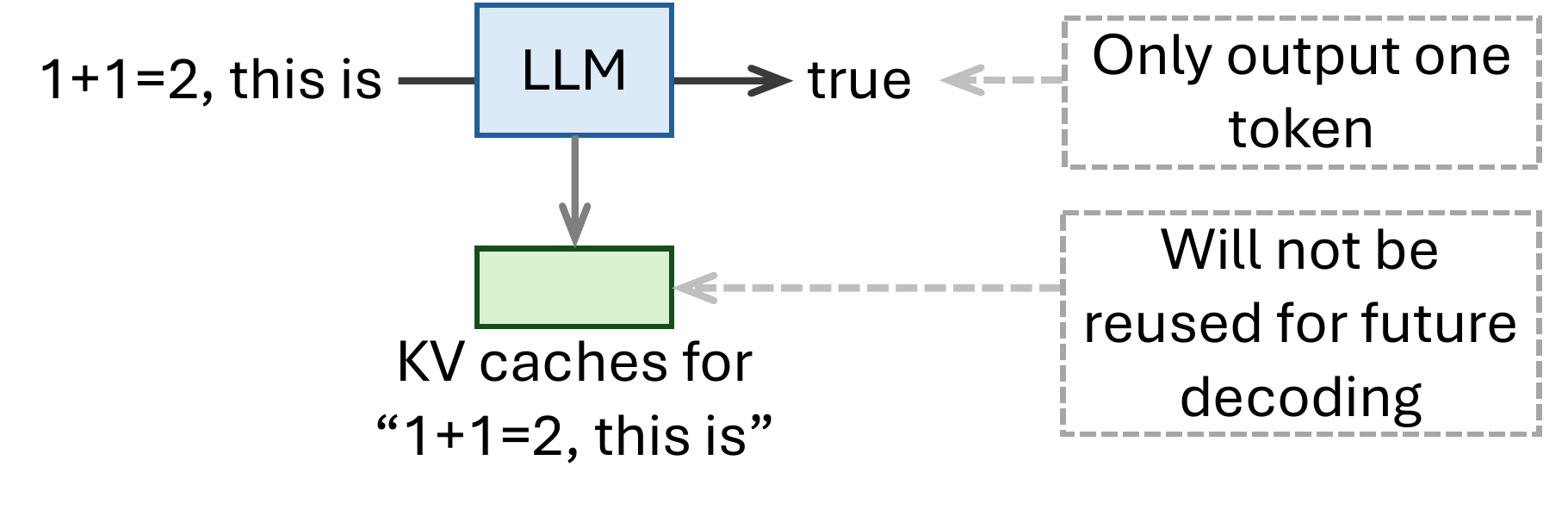}
    \subcaption{Prefill-only LLM inference}
    \end{minipage}
    
    \tightcaption{Contrasting traditional LLM inference and prefill-only LLM inference. A prefill-only request does not reuse its KV cache for long decoding as it only generates one output token.}
    \label{fig:contrast_traditional_prefill_only}
\end{figure}

\mypara{Prefill-only workloads}
Our observation is that when used in these discriminative tasks, the LLM generates only \textit{one single token as output} for each incoming request\cite{COT,rah,Doris,fingpt,SPAMT5}, because a single-token output is sufficient to express the output of these tasks, \eg a preferred choice or a label.
For example, the following prompt is used in a document recommendation application:
``\texttt{Here is the user profile: [user profile], and here is the document: [document]. Should we recommend this document to this user? Your answer is: }''.
To determine whether this document should be recommended, the LLM only needs to generate one output token (\ie \texttt{Yes} or \texttt{No}).\footnote{To ensure that LLM only generates \texttt{Yes} or \texttt{No}, the developers do not have to fine-tune the LLM or add extra prompts; instead, they can pass a list of acceptable tokens (\eg \texttt{Yes} and \texttt{No}) to the LLM engine so that LLM engine only samples output from this list.
More discussion in \S\ref{subsec:prefill-only-characteristics}.}

In this paper, we refer to such workloads as \textit{prefill-only}, because to generate one token, the LLM engine only needs to run the prefilling stage of LLM inference.

\mypara{Untapped opportunities}
Unlike traditional LLM requests, prefill-only requests present two unique opportunities.
\begin{packeditemize}
\item {\em Smaller active GPU memory footprint:} 
Normally, the LLM engine stores the KV cache of all layers in GPU memory to save repeated computation when decoding each output token. 
This amounts to a large amount of KV cache data, particularly when the input is long
(\eg the user profile in the example above may contain months of user browsing history).
For prefill-only requests, however, most stored KV cache values will not be reused, because the engine does not need to decode more than one token (as shown in Figure~\ref{fig:contrast_traditional_prefill_only}), making it possible to significantly reduce the active GPU memory footprint during the inference.
\item {\em Less uncertainty in JCT:} 
A typical LLM request's job completion time (JCT) is hard to pre-determine because the output length could be arbitrary depending on the sampling\cite{megatron,terapipe}. 
For prefill-only requests, however, the output length is always one, so the LLM engine has enough knowledge to determine the amount of work for each request, making better scheduling logic possible.
\end{packeditemize}



Unfortunately, existing LLM engines fall short of leveraging these properties of prefill-only workloads.
Because these LLM engines assume the output length of a request can be arbitrarily long, they must store in GPU memory all the KV cache of all tokens at all layers, and their request scheduling logic must handle uncertainties on JCT, rather than using the classic efficient logics like shortest-job first.

\mypara{{\em \name} fully embraces the opportunities}
This paper presents \name, the first LLM inference engine tailored for prefill-only workload.
Its technical contribution is two-fold.

First, it uses a novel {\em hybrid prefilling} to allow the LLM engine to keep only the KV cache of one layer, thus drastically reducing the active memory footprint during inference. This obviates the cross-GPU communication overheads and slowdown as each request can now fit in fewer GPUs. 
Yet, this is easier said than done. 
Naively keeping the KV cache of one layer does not directly reduce the memory footprint since the temporary tensors allocated during LLM inference still use a large amount of GPU memory, and while using chunked prefilling reduces the size of these temporary tensors, it forces the KV cache of all layers to remain in GPU memory between chunks. 

Our hybrid prefilling addresses this conundrum by processing non-attention layers chunk-by-chunk but processing attention layers normally.
The intuition is that most GPU memory usage in LLM inference comes from non-attention layers, and chunking them can significantly reduce the GPU memory footprint, as these layers are all linear layers, so each chunk can be computed independently with each other.
We note that, though hybrid prefilling does not only benefit prefill-only requests, it is the enabler for \name to potentially discard or offload the KV caches outside the GPU memory, as it ensures the prefilling is finished within one LLM forward pass rather than multiple passes that require reusing the KV caches between passes.
Also, we emphasize that hybrid prefilling merely makes it possible to not keep all the KV cache in active GPU memory during the inference --- the LLM inference engine can choose between keeping the KV cache in GPU memory, offloading part of it to CPU memory or discarding part of it.

Second, \name's request scheduling logic leverages the fact that the system has enough knowledge to pre-determine each prefill-only request's completion time. 
This makes a range of classic JCT-based scheduling algorithms possible. 
However, this again is easier said than done. 
For each incoming request, its JCT also depends on whether its prefix's KV cache is available. 
Given that the stored KV caches can constantly change, \name \textit{continuously recalibrates} the JCT of waiting requests every time before running the scheduling algorithm.
With the continuous JCT calibration, \name can more precisely determine the JCT of each request for better scheduling {\em and} maximize the chance for KV caches of a prefix to be reused across requests (as cache-hit requests have lower JCT and will be prioritized, preventing irrelevant requests from evicting the prefix cache needed by cache-hit requests). 
Moreover, for better fairness and starvation avoidance, \name also offsets the JCT based on request waiting time (as elaborated in \S\ref{subsec:continuous-jct-calibration}).

While conceptually, the prefill-only properties are not unique to LLMs but also apply to many classic deep neural networks, the exact techniques are specialized to LLMs: hybrid prefilling leverages the property that the majority of LLMs only contain attention layers and linear layers, and \name continuously calibrates the JCT based on LLM prefix KV caching.

In short, our contributions are:
\begin{packeditemize}
    \item  
    We identify the emerging trend that people use LLMs to substitute traditional deep-learning-based pipelines, and characterize a new workload -- \textit{the prefill-only workload} -- underlying such trend.
    \item We present \name, the first LLM inference engine that fully embraces the unique properties of prefill-only workloads, to drastically reduce active GPU memory footprints and enable better JCT-based scheduling. 
    
\end{packeditemize}

We implement \name on top of an enterprise-grade, state-of-the-art LLM inference engine (vLLM~\cite{vllm}).
For the generalizability of \name,  we implement hybrid prefilling via \texttt{torch.compile} so that \name can generalize to different LLM models, and implement logics related to KV cache storage without modifying hardware-related kernels so that \name generalizes to different hardware platforms.
We evaluate \name under 4 hardware setups, 3 LLM models, and 2 traces. Our evaluation shows that \name can handle upto 4$\times$ query-per-second while still achieving lower average and P99 latency than the baselines.

\section{Motivation}
\label{sec:motivation}


\subsection{Preliminary of large language models}

This section introduces the basic concepts in large language model (LLM) inference.

\mypara{Prefilling and decoding}
LLM inference contains two phases.
The prefilling phase of LLM processes the input and generates one output token.
Then, the LLM engine runs multiple rounds of the decoding phase, where each decoding phase forwards the request through the LLM and generates one more output token.

\mypara{KV caches} 
As LLM inference incurs multiple rounds, when processing one request, the LLM engine will store the intermediate tensors produced by the attention layers inside the GPU, known as KV caches, to accelerate future rounds of LLM inference.
The size of KV caches can be large---\eg the KV cache size of a request with 100,000 tokens is around 12 GB for a medium-sized LLM model (Llama-3.1-8B).

\mypara{Prefix caching} 
The KV caches generated by one request can be reused by another request if they share part of the prefix, which is commonly referred to as prefix caches.
As a result, existing LLM inference engines~\cite{PagedAttention,zheng2024sglangefficientexecutionstructured} will not immediately free these KV caches after request execution, but instead cache them in the GPU so that future requests can potentially reuse the KV caches.

\subsection{Substituting traditional deep learning models with LLMs}

In addition to generative LLM applications that generate new content for people to read and use, (\eg ChatGPT~\cite{chatgpt}, Character.AI~\cite{characterai}, GitHub Copilot~\cite{copilot}, Cursor~\cite{copilot}, Perplexity~\cite{perplexityai} and more), both industry and academia have started to use LLMs to replace traditional deep learning pipelines in applications such as recommendation, credit verification, and data labeling mainly for two reasons.

\mypara{LLM streamlines the development}
Developing traditional deep learning pipelines is time-consuming and complex, as it requires co-optimizing multiple stages in the pipeline, from data cleaning to model fine-tuning, which requires cross-team collaboration and extensive infrastructure support, and eventually accumulates maintenance debt. In contrast, LLM can take raw text as input and chat with the developer, and is general enough to obviate fine-tuning.
This allows the developer to interactively debug and improve the pipeline by just chatting with the LLM. This argument is supported by recent literature~\cite{firooz2025360brew}, which shows that a single LLM model can serve over 30 tasks across over 8 different domains. 

\mypara{LLM achieves higher decision quality}
Further, by properly selecting the size of the LLM model and engineering the prompt, LLM can generalize to out-of-domain tasks and surfaces, and achieves comparable performance similar to or better than a production model~\cite{firooz2025360brew}.

\subsection{Underlying workload: prefill-only workload}
\label{subsec:prefill-only-characteristics}

We observe that people use LLMs in these applications in a different way from those generative applications.
Concretely, these applications only let the LLM generate one single output token for each request.
We call these requests \textit{prefill-only requests} (as they only require the LLM engine to run prefilling to generate a single output token), and name such a workload a prefill-only workload.

\mypara{Single token suffices to express LLM's preference}
Generating one single token still allows LLM to express its preference between different options.
To illustrate this, we use post recommendations as an example.
Assume that we are a social media platform that aims to recommend social media posts to a specific user via LLM.
In this case, the recommendation system will first gather user's browsing history as the profile.
Then, the system selects, for example, fifty posts that the user might be interested in using traditional heuristics like embedding-based similarity search.
Then, the recommendation system sends fifty LLM requests to the LLM, one per document.
Here is an example of such LLM request:

\vspace{0.2cm}
You are a recommendation assistant to use user's profile and history to recommend the item that the user is most interested in.
Here is the user profile: 

[User profile]

Here is the browsing history of an user:

[User browsing history]

If we recommend the following article to this user, will the user be interested in reading it? Please response using Yes or No.

[Article]

Your answer is: 
\vspace{0.2cm}

Then, we constraint the output of LLM to only \texttt{Yes} and \texttt{No}, and let LLM prefill this request to yield two probability numbers: $\mathbb{P}(\texttt{Yes})$ and $\mathbb{P}(\texttt{No})$, where their sum equals to 1.
After that, the recommendation system will use $\mathbb{P}(\texttt{Yes})$ as the recommendation score.


\mypara{Lower latency}
Generating one single token also significantly reduces LLM inference latency.
This is because the input processing of LLMs is much faster  than output generation.
Concretely, using Llama 3.1-8B-model and one NVIDIA H100, we measure that handling a request with 2048 tokens input and 256 tokens output is 1.5$\times$ slower than handling a request with 2048 tokens input and 1 token output.

\mypara{Clearly-defined and controlled output behavior}
Further, we argue that prefill-only workload allows the developer to simply define and control the output behavior by passing over a list of acceptable tokens to the LLM engine and then let the LLM engine to sample only output from this list.
In contrast, it is difficult to clearly define and guarantee the expected LLM output behavior in traditional generative requests, which motivates a long line of research (\eg~\cite{xgrammar,flashinfer}).

\subsection{Characteristics of prefill-only workload}
\label{subsec:prefill-only-characteristics}

We conclude two characteristics of prefill-only workload.

First, the input length of prefill-only requests is typically long.
For example, in a documentation recommendation application, the user profile potentially contains months of the user's browsing history, which can easily reach tens of thousands of tokens, and even more.

Second, different from a traditional LLM workload that is GPU-memory-bound, a prefill-only workload is bounded by GPU computation.

\mypara{Sharing GPU resources with traditional generative workload is impractical}
To meet the stringent application requirements, instead of sharing the GPU with traditional generative requests, one needs to allocate GPU dedicated to prefill-only workloads.
This is because the inter-workload interference of LLM applications is significant.
For example, serving both prefill-only requests and traditional generative requests on the same GPUs may greatly increase the average and P99 time-per-output-token in generative use cases, as the decoding jobs will be batched with prefill jobs more frequently compared to not mixing these two workloads.

Further, the volume of prefill-only workload is large enough that it deserves dedicated GPU resources.
For example, in recommendation workload, the typical queries per second is at the scale of tens of thousands, which demands hundreds or even thousands of H100 GPUs to serve.





\subsection{Limitation of existing LLM engines}
\label{subsec:issues-of-limited-gpu-memory}

In practice, we observe two issues that limit the capacity of existing LLM inference engines in prefill-only workloads.

\mypara{Trade throughput to handle long requests}
As existing LLM engines fully store the KV caches, this limits the maximum input length (MIL in short) that they can handle, as the size of KV caches linearly scales with respect to input length. For example, the MIL is only 11,000 tokens on NVIDIA A100 40GB GPU with Qwen-32B model (FP8-quantized).

As a result, when the expected maximum request length exceeds MIL, the LLM engines have to choose one of the following approaches, and all of them come at the cost of reduced throughput.
\begin{packeditemize}
    \item \textit{Chunked prefilling}. 
    Chunked prefilling prefills the request \textit{chunk-by-chunk} so that the LLM engine can handle longer requests without parallelizing the inference.
    However, this technique reduces attention kernel performance, resulting in lower throughput.
    Concretely, we measure that chunked prefill will lower the end-to-end throughput by 14\% when chunking the input length of 20,000 with a chunk size of 512.
    Further, this technique can only marginally increase the context length by less than 2$\times$ (as shown in Table~\ref{table:maxlen_wl}), as it forces the LLM engine to fully store the KV caches of all previous chunks.
    \item \textit{Tensor parallelism}. One can increase the MIL by tensor parallelism.
    However, although tensor parallelism can reduce the latency when used together with high-speed interconnection hardware such as NVLink, it may instead inflate the latency when high-speed interconnection is not available, and it always decreases the overall throughput.
    This is because tensor parallelisms require expensive all-reduce communication between GPUs, which takes a significant portion of GPU time, even with the acceleration of NVLink and resulting in GPU computation being under-utilized during the all-reduce communication.
    \item \textit{Pipeline parallelism}. One can also increase the MIL by pipeline parallelism.
    Ideally, pipeline parallelism has the same latency-throughput trade-off as without parallelization on the same number of GPUs.
    However, pipeline parallelism introduces bubbles when the request length varies, and thus has sub-optimal latency-throughput trade-offs.
    Such bubbles can be minimized by chunking the input to the same chunk size via chunked prefilling, but chunked prefilling itself hurts the latency-throughput trade-offs as mentioned before.
\end{packeditemize}

We will further discuss parallelization-based solutions in prefix caching scenario (\S\ref{subsec:when-to-use-parallelization-instead}).


\mypara{Scheduling algorithm is unaware of JCT}
Existing LLM engines typically leverage JCT-agnostic scheduling, such as first-come-first-serve scheduling, as the output length of LLM requests can be non-deterministic and difficult to predict.
If the JCT can be accurately estimated, one can leverage JCT-aware scheduling (like shortest remaining job first) to further reduce the latency of requests.
\subsection{Opportunity and challenges}
\label{subsec:challenges}

In prefill-only workload, we characterize two optimization opportunities:
\begin{packeditemize}
    \item The active GPU memory of prefill-only inference is much lower --- the generated KV caches will not be reused for future decoding, so it can potentially be offloaded or discarded from the GPU (as shown in Figure~\ref{fig:contrast_traditional_prefill_only}).
    This allows one to potentially handle much longer requests without using solutions that degrade the throughput, like parallelizing the KV caches or chunking the input.
    \item The JCT of prefill-only requests is deterministic and predictable as the output length is always 1.
\end{packeditemize}

A naive solution to leverage the opportunities above is by offline profiling how the JCT changes with respect to request length, and in the online phase, the LLM engine uses this JCT profile to obtain the JCT of each incoming request, and then schedules the waiting requests using JCT-aware algorithms (e.g., shortest
remaining job first), and then discards the KV caches during inference.
This solution is widely adopted in traditional deep learning systems, but it has two severe limitations:

\mypara{Simply dropping KV caches increases MIL marginally}
\\
Ideally, dropping the KV caches could allow the LLM engine to handle up to the number-of-layers-time longer input length.
However, in practice, we observe that dropping the KV caches only yields 1.6$\times$ MIL improvement, measured using NVIDIA L4 using Llama-3.1-8B model.
This is because the LLM inference itself allocates large temporary tensors that consume a significant amount of GPU memory and limit MIL. See \S\ref{subsec:gpu-memory-bottleneck} for more analysis.

\mypara{Incompatible with prefix caching}
Existing LLM engines store the KV caches in the LLM engine so that future requests with the same prefix can be accelerated. This technique is known as prefix caching. Fully discarding the KV caches prevents such optimization.

\section{Overview}

In order to improve the latency-throughput trade-off in prefill-only workload, we  propose \name, the first inference engine tailored for prefill-only workload, built on top of production-level serving engine, vLLM.

\begin{figure}
    \centering
    \includegraphics[width=1.01\columnwidth]{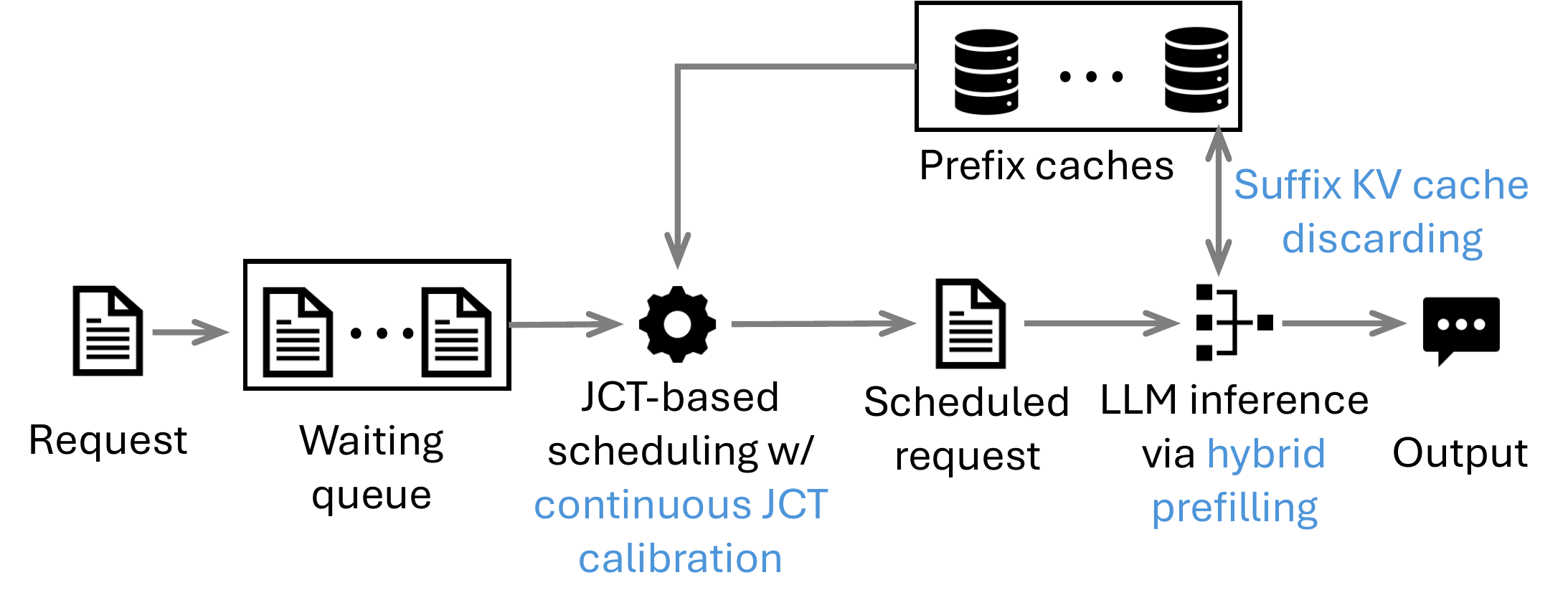}
    \vspace{-0.5cm}
    \tightcaption{Overview of \name and its core techniques.}
    \label{fig:overview}
\end{figure}

\subsection{Overview of \name}

\name is an inference engine that serves prefill-only requests online.
The workflow of \name is as follows.

\mypara{Profile run} 
\name assumes that the user provides the maximum request length (MIL) \name needs to handle.
Based on user-provided MIL, \name calculates the amount GPU memory required for LLM inference, by forwarding a fake request with maximum request length through the LLM model and measure the peak GPU memory usage during this period.
The remaining GPU memory is then used as the space for prefix KV caches. 

\mypara{During runtime}
\name opens an HTTP server compatible with the OpenAI API protocol for the user to send their prefill-only requests.
When a new request arrives, \name tokenizes the request and sends it to the waiting queue of the scheduler process using ZeroMQ-based RPC.
The scheduler process then schedules the requests in the granularity of \textit{step}.
During each step, \name enumerates the requests in the waiting queue to find the request with minimum JCT (more details in \S\ref{subsec:hybrid-prefilling-continuous-shortest-job-first}).
Then, \name pops out this request and sends it to the executor processes.
Finally, the request executor processes then execute the request (more details in \S\ref{subsec:hybrid-prefilling-continuous-shortest-job-first}) and return the prefill-only probability score all the way back to the user.

We summarize the workflow of \name in Figure~\ref{fig:overview}.

\subsection{Core techniques of \name}
\label{subsec:hybrid-prefilling-continuous-shortest-job-first}

At the core of \name is the following techniques that maximally increase MIL and throughput of prefill-only requests (we illustrate how these techniques participate in the workflow in Figure~\ref{fig:overview}):
\begin{packeditemize}
    \item \textit{Hybrid prefilling}. \name prefills non-attention layers chunk-by-chunk, but prefills the attention layers normally.
    This technique allows \name to maximally increase the MIL by simultaneously chunking the intermediate tensors of non-attention layers and discarding KV caches for attention layers, while not hurting the performance of attention operations.
    \item \textit{Suffix KV cache discarding / offloading}.
    \name discards / offloads the KV caches of suffix tokens when the KV caches cannot fit into the GPU, allowing \name to handle much longer requests without using solutions (chunked prefill or parallelizing the inference) that create extra overhead and degrade throughput. 
    In this paper, for simplicity, \name discards the useless KV cache, though in practice \name can also choose to offload the KV cache to CPU memory if such an offloading mechanism is available.
    \item \textit{Continuous JCT calibration}. \name continuously re-estimates the JCT of each request based on what requests are previously scheduled, and then schedules just one request with the lowest JCT.
    This allows \name to obtain accurate JCT estimation using what requests are previously scheduled, which allows \name to increase the prefix cache hit rate by dynamically identifying the requests that can hit the cache created by the newly-scheduled requests, and thus reduces the request latency.
\end{packeditemize}

\section{Hybrid prefilling}

In this section, we first introduce hybrid prefilling, the key optimization for \name to improve MIL by maximally controlling the GPU memory footprint during LLM inference.

\subsection{Bottleneck: intermediate tensors of linear layers}
\label{subsec:gpu-memory-bottleneck}

As mentioned in \S\ref{subsec:issues-of-limited-gpu-memory}, simple solutions only marginally increase MIL: storing the KV cache of active layer increases MIL by merely 1.6$\times$, while chunked prefilling increases MIL by 2$\times$ but at the cost of reducing the kernel performance of attention operation, thus reducing throughput.

To understand why simply discarding the KV cache of active layers does not work well, we profile how the GPU memory usage of PyTorch GPU memory allocator varies over time when prefilling a request with $32,768$ tokens, using the Llama-3.1-8B model.

\begin{figure}[htbp]
    \centering
    \begin{minipage}[b]{0.48\textwidth}
        \noindent
        \centering
        \includegraphics[width=\linewidth]{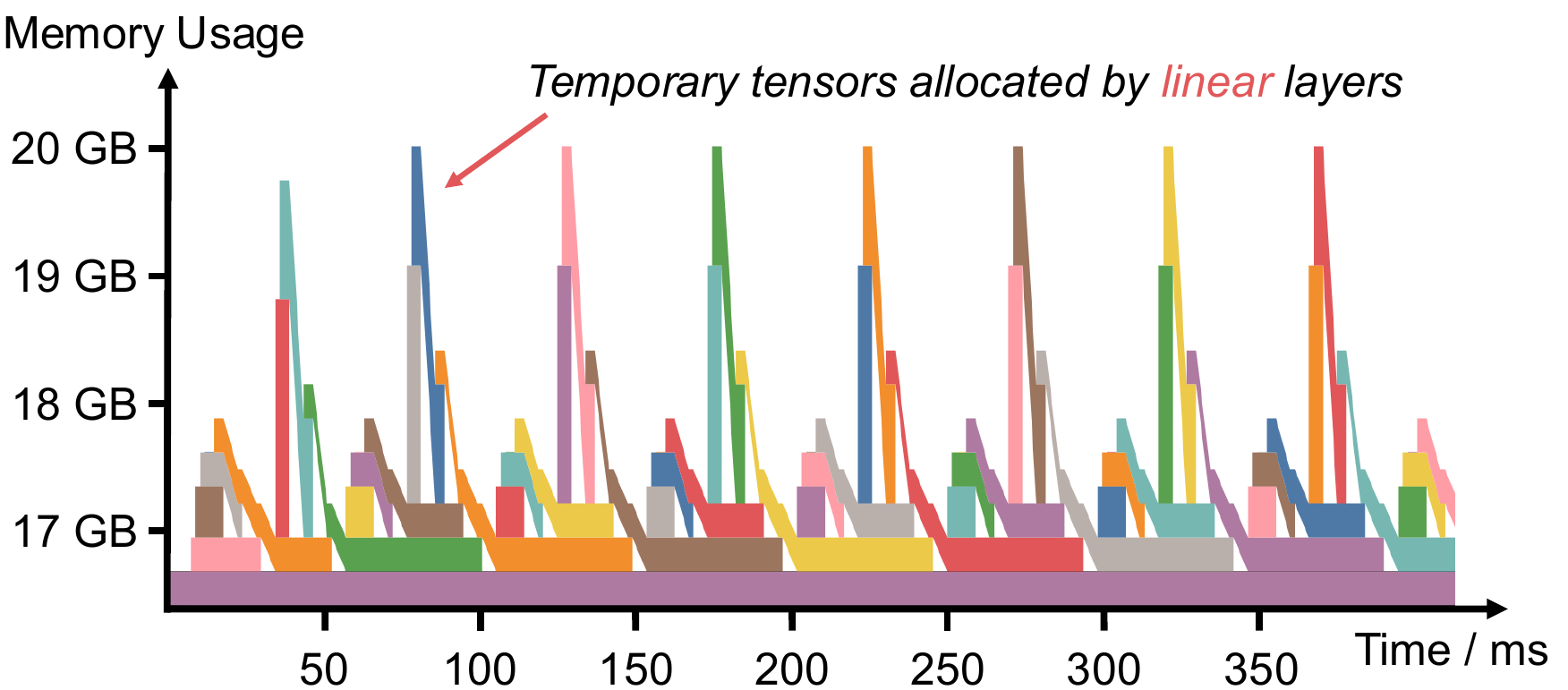}
        
        \subcaption{Without hybrid prefilling}
        \label{fig:memory_usage_vs_time_baseline}
    \end{minipage}

    \hfill
    \begin{minipage}[b]{0.48\textwidth}
        \noindent
        \centering
        \includegraphics[width=\linewidth]{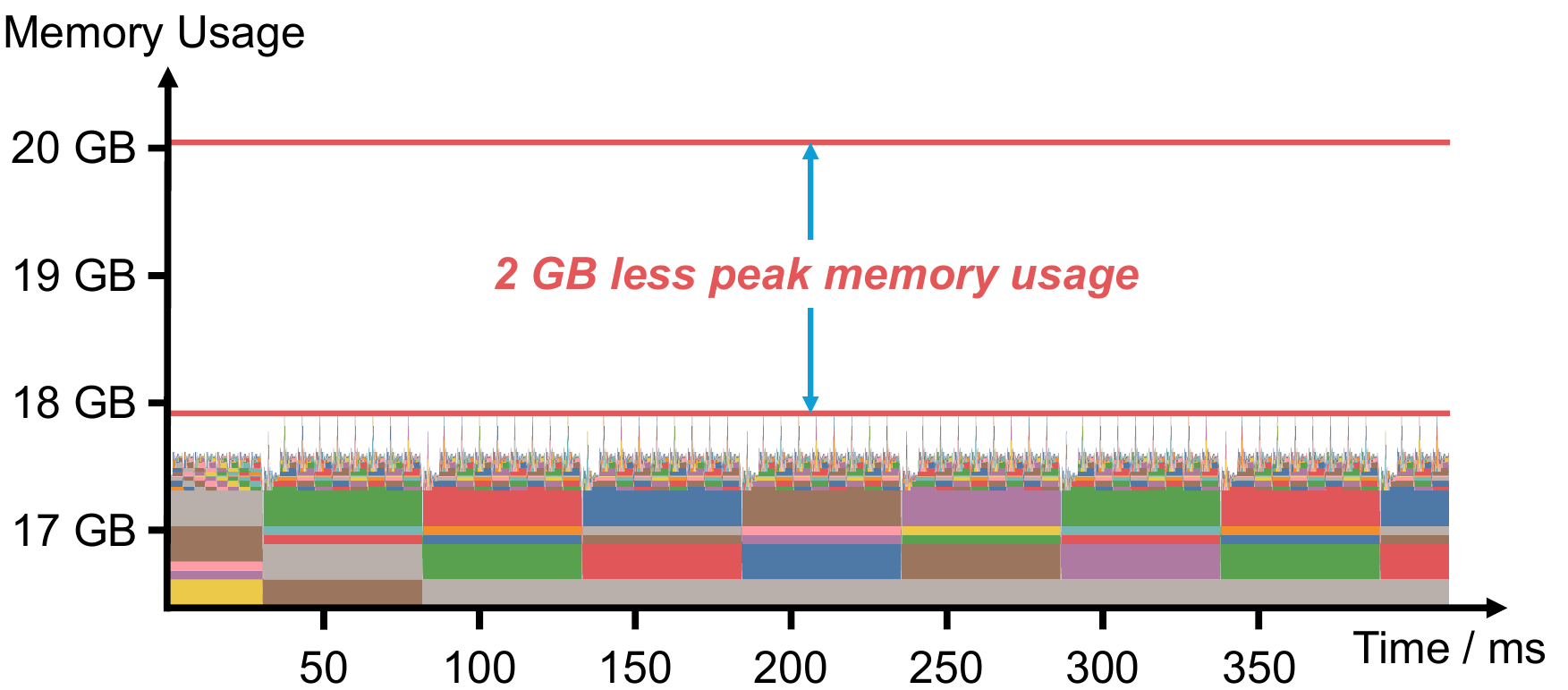}
        
        \subcaption{With hybrid prefilling}
        \label{fig:memory_usage_vs_time_optimized}
    \end{minipage}
    \tightcaption{GPU memory traces of prefilling 32,768 tokens through Llama-3.1-8B model. intermediate tensors allocated by linear layers create large GPU memory spikes and significantly increase peak GPU memory usage of prefilling, while hybrid prefilling forwards the non-attention layers chunk-by-chunk and thus reduces the peak GPU memory usage.} 
    \label{fig:memory_usage_vs_time_illustrative}
\end{figure}

\begin{figure}[tbp]
    \centering
    \includegraphics[width=1.0\linewidth]{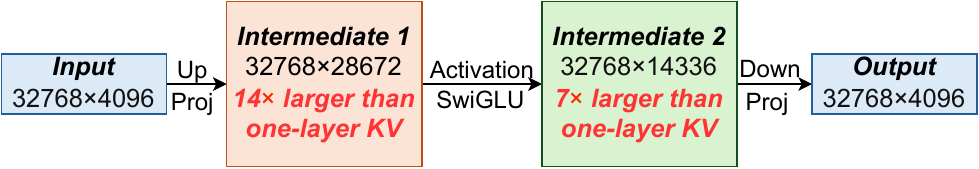}
    \tightcaption{Illustrating the tensor sizes of the MLP module in Llama-3.1-8B with bfloat16 precision. The size of intermediate tensors is much larger than the size of one-layer KV cache.}
    \label{fig:intermediate-tensors}
\end{figure}

As shown in Figure~\ref{fig:memory_usage_vs_time_baseline}, we see that there are periodic spikes during the prefilling of the request.
These spikes correspond to the intermediate tensors allocated to store the input and output of MLP module (a sequence of linear layers) inside Llama models.
These tensors are large, as they contain 28672 floating numbers per token,  14$\times$ larger than the KV cache size of one layer. To illustrate why the intermediate tensors are large, we visualize the forward pass of the MLP module in the Llama-3.1-8B model in Figure~\ref{fig:intermediate-tensors}, along with each intermediate tensor's shape and GPU memory occupation in bfloat16 precision.

We note that this trend---intermediate tensors allocated by linear layers are much larger than the KV cache size of one layer---also applies to existing LLM models.
This is because existing LLM models choose to lower the size of KV cache (in order to increase the decoding throughput by enlarging the batch size, and also to handle longer LLM requests) without sacrificing the number of parameters in the LLM model.
As a result, the LLM models have to inflate the tensor in MLP module, so that one can incorporate sufficient parameters into the linear layers.



\subsection{Controlling intermediate tensors via hybrid prefilling}

To control the size of intermediate tensors, we propose a new technique: \textit{hybrid prefilling}, where we prefill the non-attention layers chunk-by-chunk and prefill the attention layers normally.
Such hybrid prefilling will not change the LLM inference results, as the non-attention layers in LLMs are linear layers, and thus each chunk can be calculated separately from the other.


To show why hybrid prefilling reduces the GPU memory footprint of non-attention layers, we contrast traditional prefilling and hybrid prefilling in Figure~\ref{fig:memory_usage_vs_time_illustrative}.
We see that, for traditional prefilling, the GPU memory trace contains high peaks because it stores the intermediate tensors for all input tokens, while in hybrid prefilling, we only store the intermediate tensor for just one chunk at any given point of time and thus the peak GPU memory usage is significantly reduced compared to traditional prefilling.




\subsection{Implement hybrid prefilling via \texttt{torch.compile}}


We employ \texttt{torch.compile} to implement hybrid prefilling. \texttt{torch.compile} is a new feature in PyTorch that allows the developer to change computation graph of one model without touching the model inference code.

We implement hybrid prefilling on top of the computation graph compiled by \texttt{torch.compile}.
Concretely, we group the consecutive linear operations into a large virtual layer.
Then, we forward through such large virtual layer chunk-by-chunk, and concatenate the output tensors of each chunk at the end.

We clarify some detailed optimizations to the hybrid prefilling process:

\begin{packeditemize}
    \item \textit{Output preallocation}: The above process requires concatenating the output tensor chunks to a large tensor, which doubles the GPU memory footprint of output tensor.
    To avoid this, we preallocate the output tensor before the forward pass using shape information inferred from the computational graph, 
    and directly write the output tensor for each chunk into this pre-allocated output tensor.
    \item \textit{In-place computation}: We reuse the GPU memory of the input tensor to store the output tensor if they have the same shape.
    This is because the relative position between output chunk $i$ and the output tensor is exactly the same as those between input chunk and the input tensor.
\end{packeditemize}

We evaluate the effectiveness of these optimizations in \S\ref{sec:evaluation}.

\section{Enabling prefix caching}

In this section, we illustrate how we enable prefix caching in \name while still preserving the benefit of discarding the KV caches, and how \name improves the scheduling algorithm in the context of prefix caching.

\subsection{Suffix KV cache discarding}

To simultaneously enable prefix caching and allow \name to increase MIL by dropping KV caches, we propose suffix KV cache discarding, where \name maximally preserves the KV cache of prefix tokens in GPU and discards the KV cache of suffix tokens.

We note that hybrid prefilling is the enabler of this technique, as hybrid prefilling only prefills each request within one LLM inference, allowing one to potentially discard part of the KV cache without worrying about slowing down the inference.

Also, our implementation does not change the hardware kernels as we reuse the abstractions created by sliding window attention in vLLM~\cite{vllm-code-repo} to implement suffix KV cache discarding.

\subsection{Comparing with parallelization}
\label{subsec:when-to-use-parallelization-instead}

\mypara{Dropping KV caches improves throughput when no prefix cache reusing}
Compared to other approaches that increase MIL, \name can process requests at the highest throughput as it does not parallelize the LLM inference or chunk the attention operation. As a result, when there is no prefix cache reusing, one should use \name to maximally improve the throughput.

\mypara{Parallelization can lower latency under high-speed inter-GPU interconnection}
However, tensor parallelization can improve the latency of the LLM serving system under low QPS at the cost of extra communication.
As a result, if the GPUs are interconnected with high-speed hardware like NVLink, one can use tensor parallelization to maximally reduce the latency.

\section{Scheduling in the context of prefix caching}

In this section, we discuss the scheduling algorithm of \name in the prefix caching context. 
We first discuss why \name does not choose to batch prefill-only requests.
Then, we characterize two requirements for the scheduling algorithm in the context of prefix caching, and propose continuous JCT calibration as a general component for JCT-aware scheduling to meet the two requirements.

\subsection{Why not batching prefill-only requests}

In traditional LLM workloads, to maximize the output generation throughput (\textit{i.e.} decoding throughput), the inference engine needs to maintain a large batch size by continuously batching new requests to the LLM engine and removing finalized requests from the batch.
This technique significantly improves throughput, because the output generation phase is bounded by GPU memory accessing bandwidth, and batching 2$\times$ requests only marginally increases the total GPU memory needs to be accessed (as the major part is the LLM model weights) and thus marginally increases the runtime, but doubles the generation throughput.

However, as shown in \S\ref{subsec:prefill-only-characteristics}, the inference in prefill-only workload is typically GPU computation-bound, and thus, batching does not significantly improve the throughput of processing
Thus, batching prefill-only requests increases the average latency compared to processing the requests one by one, and does not improve the throughput.
As a result, \name chooses to schedule the requests one by one instead of batching them.

\subsection{Limitation of traditional JCT-based scheduling}
\label{subsec:traditiona-jct-scheduling-limitation}

We observe that, traditional JCT-based scheduling algorithm has a low prefix cache hit rate in the context of prefix caching, resulting in high latency and low throughput.

This is because the JCT changes over time when prefix caching is enabled: the JCT of one request reduces when the prefix cache related to this request enters the LLM engine, and increases when this prefix cache is evicted.
As a result, this approach fails to timely prioritize those requests that can hit the prefix cache, and when the LLM engine executes these requests, the prefix cache might already be evicted.

This is because traditional JCT-based scheduling algorithms make decisions based on the JCT when the request arrives, and thus fail to timely prioritize the requests when the LLM engine receives prefix cache related to these requests, and deprioritize the requests when the corresponding prefix cache of these requests is evicted.

\begin{figure}
    \centering
    \includegraphics[width=0.98\columnwidth]{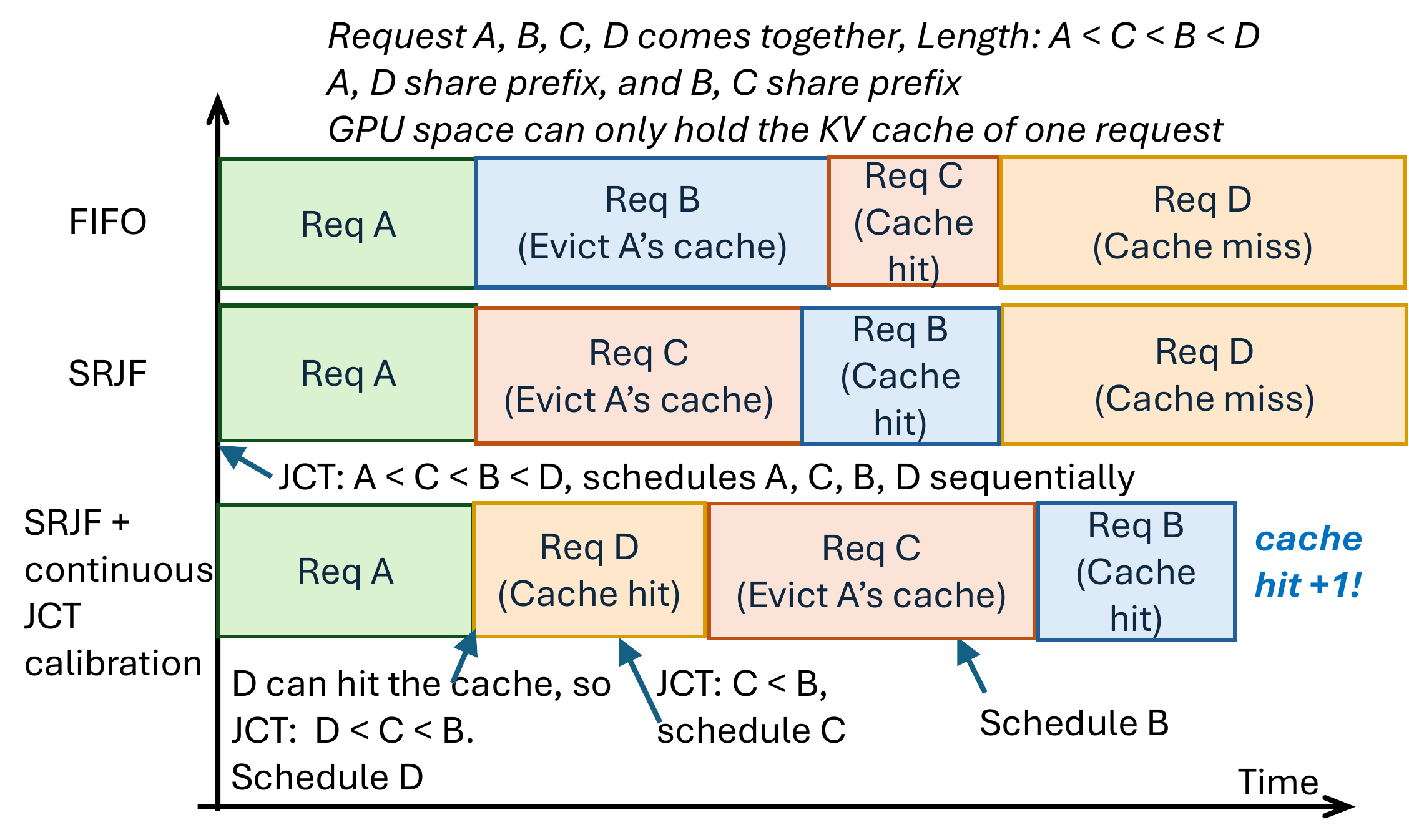}
    \tightcaption{Contrasting first-in-first-out (FIFO) scheduling, shortest-remaining-job-first (SRJF) scheduling and \name's SRJF scheduling with continuous JCT calibration. The scheduling of \name yields one more cache hit, achieving lower average latency.}
    \label{fig:scheduling-difference-illustrative-example}
\end{figure}

\mypara{Example}
To further understand why the traditional JCT-based scheduling algorithm has low prefix cache hit rates. We provide an illustrative example.
In this example, we assume that four requests (A, B, C, D) come into the LLM inference engine altogether, with request length A < C < B < D.
Further, we assume A and D share the same prefix, and so do B and C.
Also, we assume that the prefix cache space is limited and thus can only store the state of one request.

In this case, traditional shortest remaining job first scheduling will schedule the job based on its JCT, which is proportional to the request length. 
As a result, it will schedule the request in the order of A, C, B, D.
In this case, request B can hit the prefix cache of request C, and thus request B can be accelerated.
However, request D that could have hit the prefix cache of A, cannot be accelerated (as request C evicted the state of A), leading to high inference latency.

\subsection{Continuous JCT calibration}
\label{subsec:continuous-jct-calibration}

\begin{algorithm}
\caption{SRJF with continuous JCT calibration}
\label{algo:shortest-remaining-job-first}
\begin{algorithmic}[1]
\State \textbf{Input:} Waiting queue $Q$ of requests
\State \textbf{Output:} Request to schedule in the next step

\State $r_{\text{shortest}} \gets \text{None}$
\State $score_{\text{min}} \gets \infty$

\For{each request $r \in Q$}
    \State $n_{\text{input}} \gets$ the number of input tokens in $r$
    \State $n_{\text{cached}} \gets$ the number of tokens in $r$ that hits prefix cache
    \State $T_{\text{queue}} \gets$ the queuing time of $r$
    \State $score \gets \text{get\_jct}(n_{\text{input}}, n_{\text{cached}}) - \lambda \cdot T_{queue}$
    \If{$score < score_{\text{min}}$}
        \State $score_{\text{min}} \gets jct$
        \State $r_{\text{shortest}} \gets r$
    \EndIf
\EndFor

\State \textbf{Schedule} $r_{\text{shortest}}$
\end{algorithmic}
\end{algorithm}

To improve the prefix cache hit rate of JCT-based scheduling algorithm, we propose  \name employs \textit{continuous JCT calibrates}, where \name calibrates the JCT of waiting requests every time before scheduling.

Such calibration significantly improves the prefix cache hit rate, as it allows the scheduling algorithm to timely prioritize those requests that can hit the prefix cache (as they typically have much lower JCT than other requests), which makes these requests much more likely to hit the prefix cache.

\mypara{Example}
To further illustrate why continuous JCT calibration helps JCT-based scheduling algorithms, like SRJF, improve latency, we give an illustrative example using the same setup as \S\ref{subsec:traditiona-jct-scheduling-limitation}.
The first scheduled job will be A.
When scheduling the next job, the scheduler will perform JCT calibration, where it finds out that the JCT of D is significantly lowered as D can hit the prefix cache of A.
As a result, the second scheduled job will be D.
Then, the scheduler calibrates the JCT of B and C again, and their JCT remains unchanged.
As a result, the scheduler then schedules C as it is shorter (and thus has lower JCT).
After that, the scheduler will then schedule request B.
In this case, the total number of cache hits is 2, where the total number of cache hits is 1 in both FIFO scheduling and naive SRJF scheduling.
Thus, SRJF with continuous calibration achieves lower latency and higher throughput.

\mypara{Calibration details}
As shown in Algorithm~\ref{algo:shortest-remaining-job-first}, \name calibrates the JCT of a given request $r$ by calculating the number of input tokens $n_{\text{input}}$ and the number of tokens that hits the prefix cache $n_{\text{cached}}$, and generate the JCT of this request by calling $jct(n_{\text{input}}, n_{\text{cached}})$, where we obtain $jct$ by profiling how the JCT varies with respect to different pairs of $n_{\text{input}}$ and $n_{\text{cached}}$ that covers the maximum input length with the granularity of 1000 tokens, and trains a small linear model using linear regression.

Empirically, however, we found that the number of tokens that do not hit the prefix cache (\ie $n_{\text{input}} - n_{\text{cached}}$) is a good proxy of JCT: we measure that, on 1$\times$ A100 the Pearson correlation coefficient between the actual JCT and the number of cache miss tokens $n_{\text{input}} - n_{\text{cached}}$ is 0.987 on Qwen 32B with FP8 quantization (where 1 means perfectly correlated).
As a result, \name uses this JCT proxy by default.

\mypara{Preventing starvation}
In order to prevent starvation, \name will reduce the JCT by $\lambda \cdot T_{req}$, where $T_{req}$ is the queuing time of the request and $\lambda$ is a hyperparameter: increasing the $\lambda$ result in better worst-case latency at the trade of worse average latency.

We summarize the scheduling algorithm in Algorithm \ref{algo:shortest-remaining-job-first}.







\tightsection{Evaluation}
\label{sec:evaluation}

\begin{table*}[t]
\footnotesize
\begin{tabular}{lllllll}
\hline
\hline
Dataset                                                       & Why evaluating this dataset                                                                  & \# of users & User profile length                                                       & Post length & \# of requests per user                                                                                 & Total \# of tokens \\
\hline
\begin{tabular}[c]{@{}l@{}}Post\\ recommendation\end{tabular} & \begin{tabular}[c]{@{}l@{}}Evaluate the ability of \name\\under frequent prefix cache reuse\end{tabular}    & 20          & \begin{tabular}[c]{@{}l@{}}11, 000 tokens\\--- 17, 000 tokens\end{tabular} & 150 tokens  & \begin{tabular}[c]{@{}l@{}}50\\(Each request corresponds\\to one post)\end{tabular} & 14,000,000       \\
\hline
\begin{tabular}[c]{@{}l@{}}Credit\\ verification\end{tabular} & \begin{tabular}[c]{@{}l@{}}Evaluate the ability of \name\\under long input length\end{tabular} & 60          & \begin{tabular}[c]{@{}l@{}}40, 000 tokens\\--- 60, 000 tokens\end{tabular} & N/A         & 1                                                                                                       & 3,000,000       \\
\hline
\hline
\end{tabular}
\vspace{0.3cm}
\tightcaption{Summarizing the dataset used in the evaluation of \name and baselines.}
\label{table:dataset}
\end{table*}

\begin{table}[t]
\footnotesize
\centering
\begin{tabular}{l|ccc}
\hline
\hline
\textbf{ Config} & \multicolumn{3}{c}{\textbf{Max input length (measured by number of tokens)}} \\
\cline{2-4}
 & \textbf{L4} & \textbf{A100} & \textbf{H100} \\
\hline
\begin{tabular}[c]{@{}c@{}}Paged\\Attention\end{tabular} & 
\begin{tabular}[c]{@{}c@{}}$24,000$\\ WL1: $\checkmark$, WL2: $\times$\end{tabular} &
\begin{tabular}[c]{@{}c@{}}$11,000$  \\ WL1: $\times$, WL2: $\times$\end{tabular} &
\begin{tabular}[c]{@{}c@{}}$15,000$\\ WL1: $\times$, WL2: $\times$\end{tabular} \\
\hline
\begin{tabular}[c]{@{}c@{}}Chunked\\Prefill\end{tabular} & 
\begin{tabular}[c]{@{}c@{}}$46,000$\\ WL1: $\checkmark$, WL2: $\times$\end{tabular} &
\begin{tabular}[c]{@{}c@{}}$17,000$ \\ WL1: $\checkmark$, WL2: $\times$\end{tabular} &
\begin{tabular}[c]{@{}c@{}}$25,000$\\ WL1: $\checkmark$, WL2: $\times$\end{tabular} \\
\hline
\begin{tabular}[c]{@{}c@{}}Pipeline\\Parallel\end{tabular} & 
\begin{tabular}[c]{@{}c@{}}$72,000$\\ WL1: $\checkmark$, WL2: $\checkmark$\end{tabular} &
\begin{tabular}[c]{@{}c@{}}$38,000$ \\ WL1: $\checkmark$, WL2: $\checkmark$\end{tabular} &
\begin{tabular}[c]{@{}c@{}}$183,000$\\ WL1: $\checkmark$, WL2: $\checkmark$\end{tabular} \\
\hline
\begin{tabular}[c]{@{}c@{}}Tensor\\Parallel\end{tabular} & 
\begin{tabular}[c]{@{}c@{}}$195,000$\\ WL1: $\checkmark$, WL2: $\checkmark$\end{tabular} &
\begin{tabular}[c]{@{}c@{}}$77,000$ \\ WL1: $\checkmark$, WL2: $\checkmark$\end{tabular} &
\begin{tabular}[c]{@{}c@{}}$238,000$\\ WL1: $\checkmark$, WL2: $\checkmark$\end{tabular} \\
\hline
\begin{tabular}[c]{@{}c@{}}\name\\(ours)\end{tabular} & 
\begin{tabular}[c]{@{}c@{}}$130,000$\\ WL1: $\checkmark$, WL2: $\checkmark$\end{tabular} &
\begin{tabular}[c]{@{}c@{}}$87,000$ \\ WL1: $\checkmark$, WL2: $\checkmark$\end{tabular} &
\begin{tabular}[c]{@{}c@{}}$97,000$\\ WL1: $\checkmark$, WL2: $\checkmark$\end{tabular} \\
\hline
\hline
\end{tabular}
\vspace{0.3cm}
\tightcaption{Evaluating the max input length that \name and baselines can handle under various hardware setups.
WL1 indicates the post recommendation workload, and WL2 indicates the credit verification workload. $\times$ means that the max input length is insufficient to run corresponding workload.}
\label{table:maxlen_wl}
\end{table}

Our evaluation shows that:
\begin{packeditemize}
    \item \name handles $1.4 - 4.0\times$ larger query-per-second without inflating the average latency and P99 latency compared to baselines.
    \item \name expands the maximum request length by upto 5$\times$ without requiring parallelizing the LLM inference.
\end{packeditemize}

\subsection{Evaluation setup}

\begin{figure*}[t]
    \centering
    \includegraphics[width=0.77\linewidth]{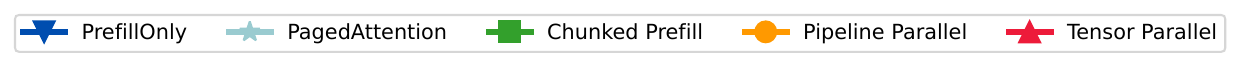}
    \vspace{0.5em}
    \begin{minipage}{0.24\textwidth}
        \centering
        \includegraphics[width=\linewidth]{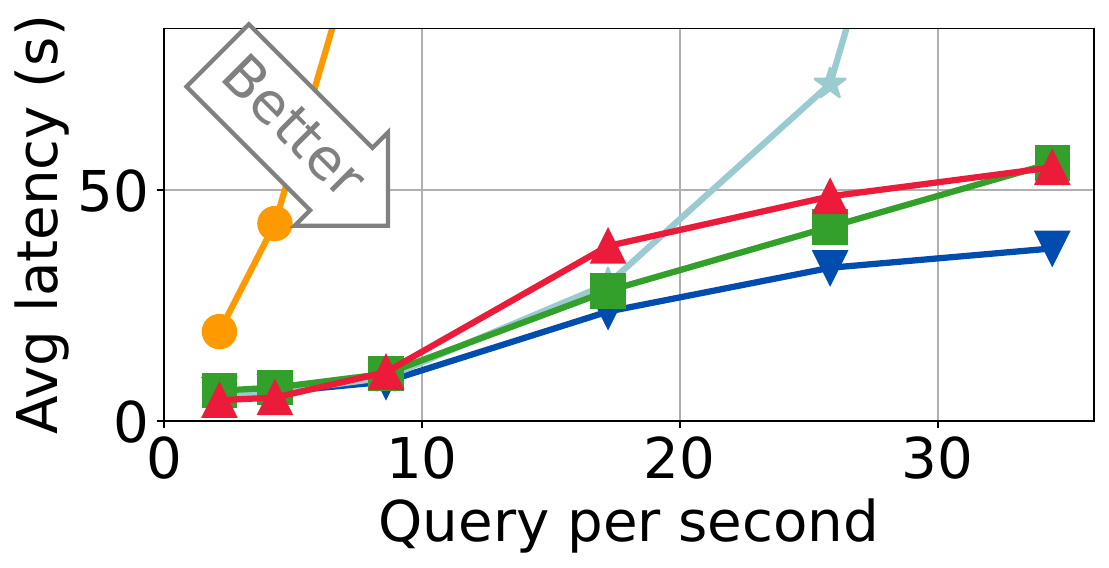}
        \subcaption{Post recommendation\\L4\\ QPS --- mean latency}
    \end{minipage}
    \begin{minipage}{0.24\textwidth}
        \centering
        \includegraphics[width=\linewidth]{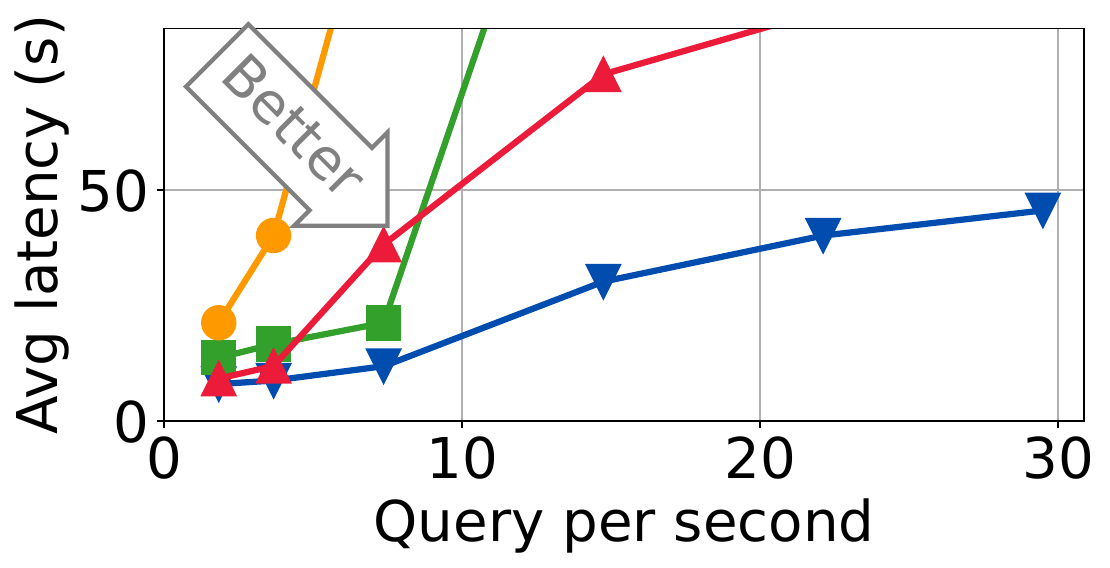}
        \subcaption{Post recommendation\\A100\\ QPS --- mean latency}
    \end{minipage}
    \begin{minipage}{0.24\textwidth}
        \centering
        \includegraphics[width=\linewidth]{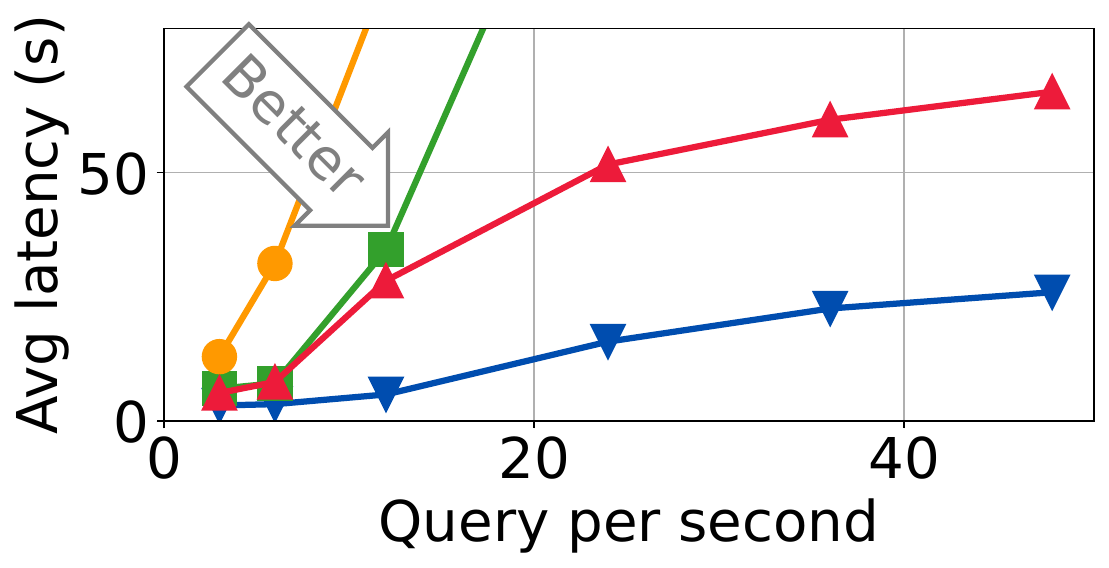}
        \subcaption{Post recommendation\\H100 w/o NVLink\\ QPS --- mean latency}
    \end{minipage}
    \begin{minipage}{0.24\textwidth}
        \centering
        \includegraphics[width=\linewidth]{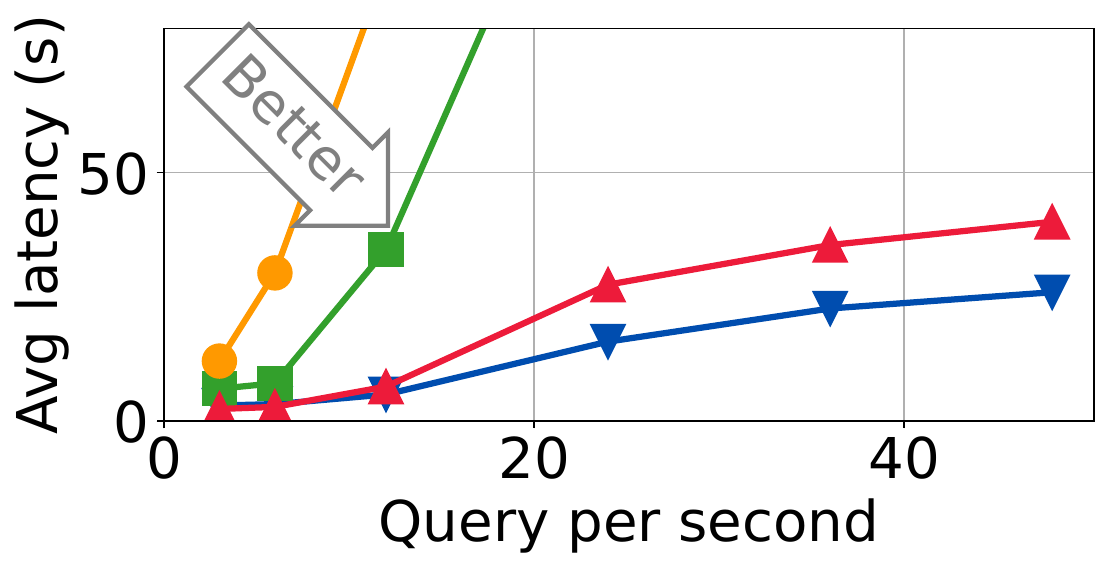}
        \subcaption{Post recommendation\\H100 w/ NVLink\\ QPS --- mean latency}
    \end{minipage}

    \vspace{0.5em}

    \begin{minipage}{0.24\textwidth}
        \centering
        \includegraphics[width=\linewidth]{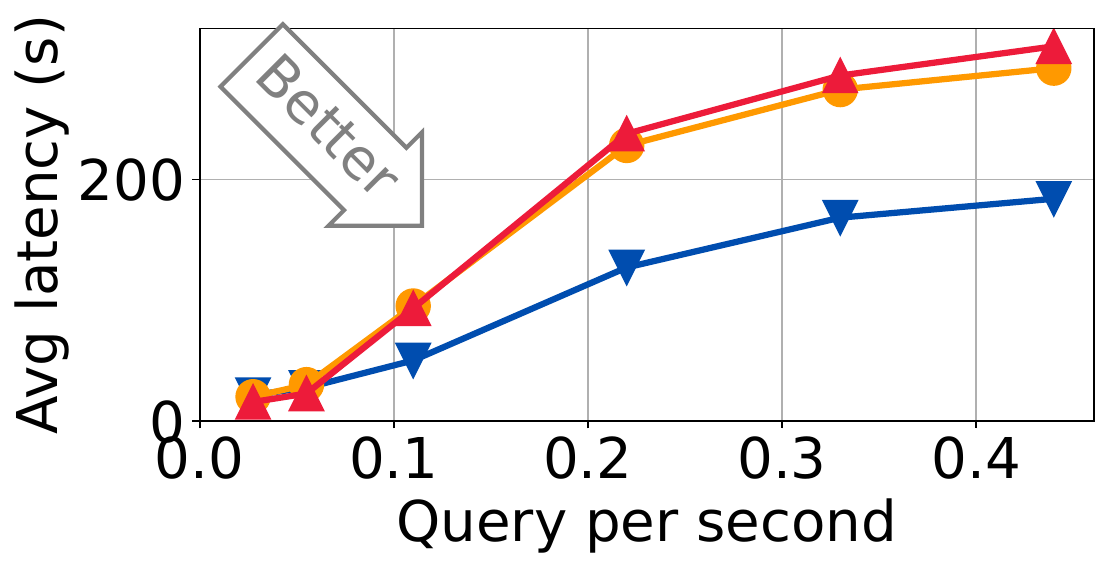}
        \subcaption{Credit verification\\L4\\ QPS --- mean latency}
    \end{minipage}
    \begin{minipage}{0.24\textwidth}
        \centering
        \includegraphics[width=\linewidth]{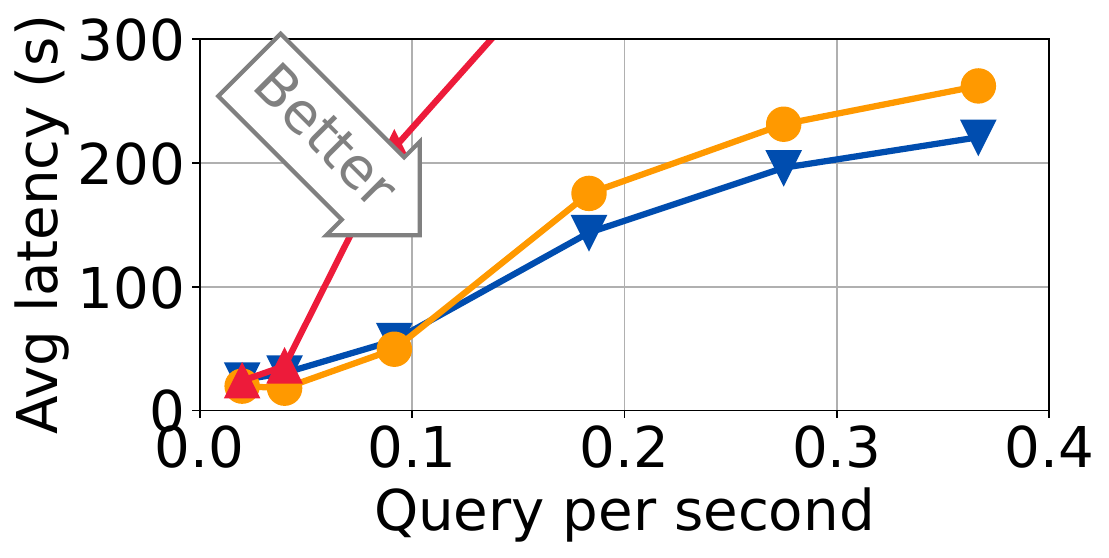}
        \subcaption{Credit verification\\A100\\ QPS --- mean latency}
    \end{minipage}
    \begin{minipage}{0.24\textwidth}
        \centering
        \includegraphics[width=\linewidth]{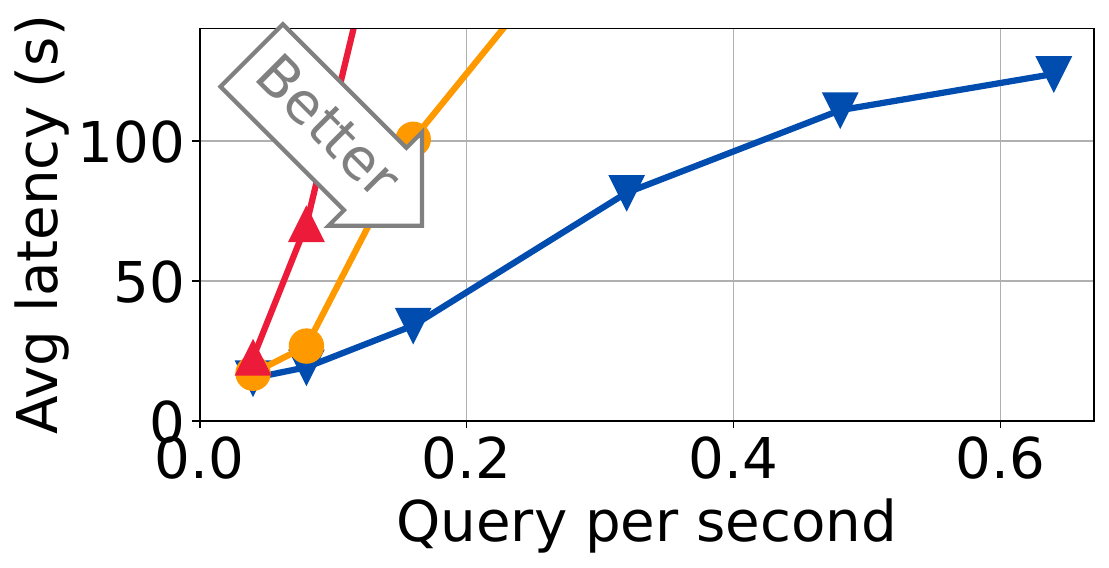}
        \subcaption{Credit verification\\H100 w/o NVLink\\ QPS --- mean latency}
        \label{fig:latency-wl2-h100}
    \end{minipage}
    \begin{minipage}{0.24\textwidth}
        \centering
        \includegraphics[width=\linewidth]{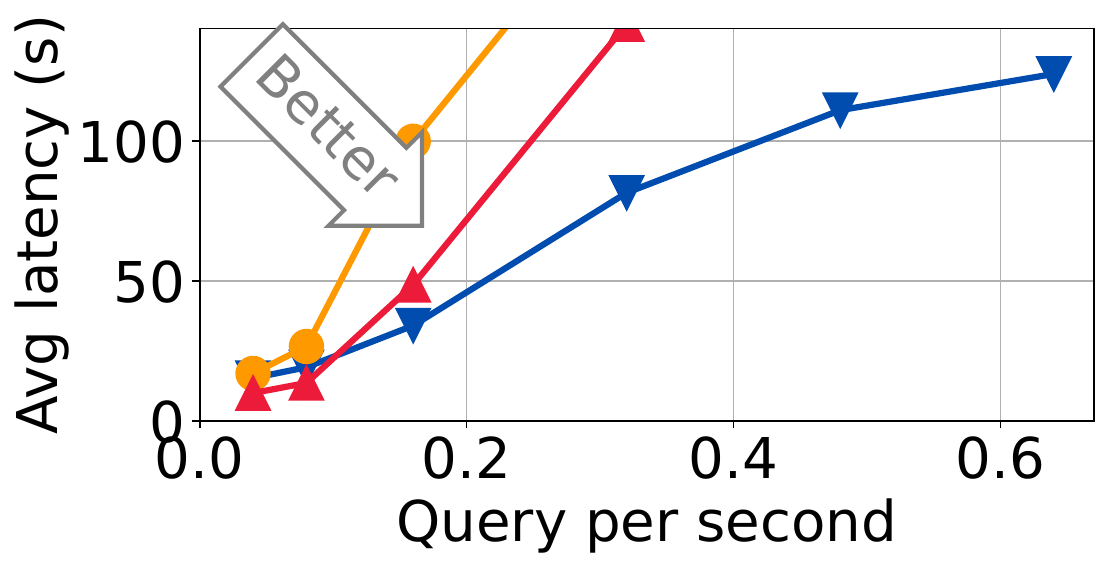}
        \subcaption{Credit verification\\H100 w/ NVLink\\ QPS --- mean latency}
        \label{fig:latency-wl2-h100-nvlink}
    \end{minipage}

    \tightcaption{QPS --- mean latency trade-off of \name and baselines on four different hardware setups and two applications.
    \name significantly reduces the latency when the QPS is high and only has higher QPS than tensor parallel baseline when QPS is low.
    Though tensor parallelism sometimes have lower latency than \name under low QPS, it has much lower throughput than \name due to extra communication cost and thus scales much worse than \name in high QPS.
    }
    \label{fig:overall-evaluation-mean-latency}
\end{figure*}

In this subsection, we introduce the dataset, GPUs, LLM models, evaluation metrics, and baselines in our evaluation.

Existing LLM datasets mainly focus on evaluating the LLM accuracy instead of the performance of the LLM engine.
As a result, in our evaluation, we use two simulated datasets, covering two tasks: post recommendation on a social media platform and credit verification for a bank application. 

We summarize our evaluation dataset in Table~\ref{table:dataset}. 

\mypara{Post recommendation dataset}
In this dataset, we aim to evaluate the benefit of \name under a short context scenario, where the major benefit of \name comes from its scheduling.
Concretely, we simulate a post recommendation scenario, where we recommend 10 out of 50 posts for a given user, based on the browsing history of this user. 
The key attributes from the perspective of LLM engine performance are the following parameters:
\begin{packeditemize}
    \item \textit{Post length:} To get a rough estimation of the post length, in terms of number of tokens, we take X as an example, and measured that the number of tokens for a short X post is less than 150 tokens.
    As a result, we use 150 tokens as the post length.
    \item \textit{Number of posts to be recommended per user}: We set the number of posts to be recommended per user as 50. We assume that these 50 posts are given by the underlying recommendation systems using heuristics like embedding-based similarity search.
    \item \textit{User profile length}: 
    As for the user profile length, we focus on the click history of one user, where the user has already been engaged with the social media four times a week, and for four weeks.
    We assume that each time the user only clicks on five or six posts.
    As a result, the total length of the profile history is roughly 11,000 to 17,000 tokens.
    As a result, we use a normal distribution to simulate the user profile length, with a mean as 14,000 and a standard deviation of 3,000.
    \item \textit{Number of users}: We evaluated 20 users in total.
\end{packeditemize}

\begin{figure*}[t]
    \centering
    \includegraphics[width=0.77\linewidth]{figures/eval_figures/legends.pdf}
    \vspace{0.5em}
    \begin{minipage}{0.24\textwidth}
        \centering
        \includegraphics[width=\linewidth]{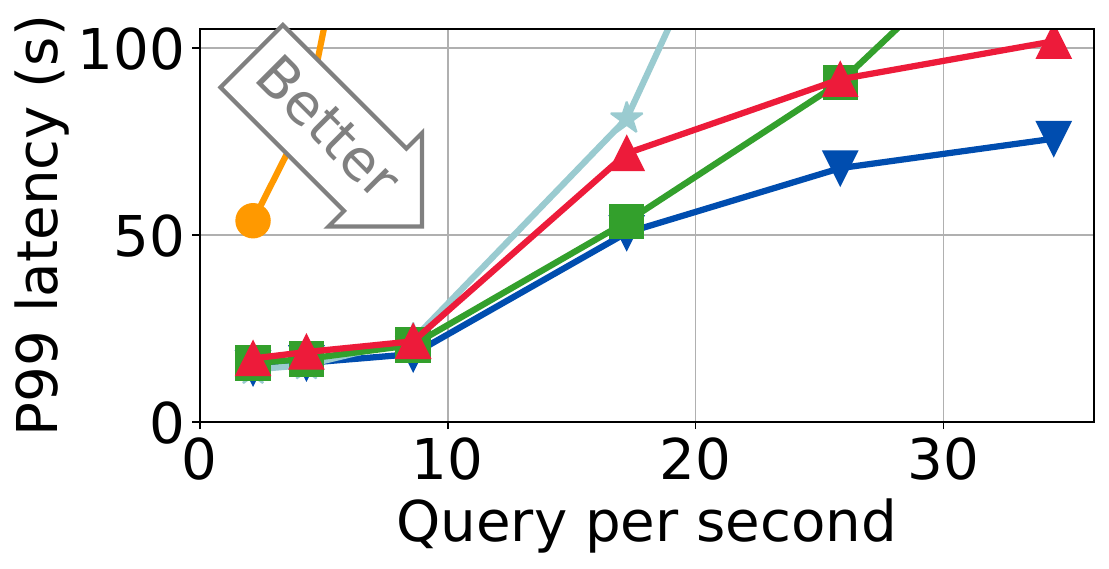}
        \subcaption{Post recommendation\\L4\\ QPS --- p99 latency}
    \end{minipage}
    \begin{minipage}{0.24\textwidth}
        \centering
        \includegraphics[width=\linewidth]{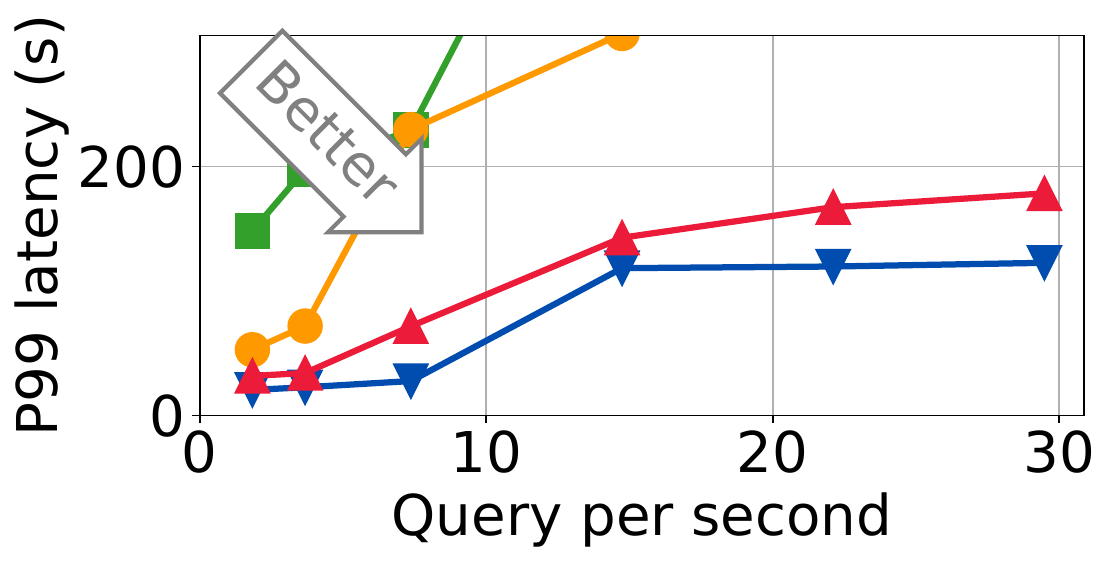}
        \subcaption{Post recommendation\\A100\\ QPS --- p99 latency}
    \end{minipage}
    \begin{minipage}{0.24\textwidth}
        \centering
        \includegraphics[width=\linewidth]{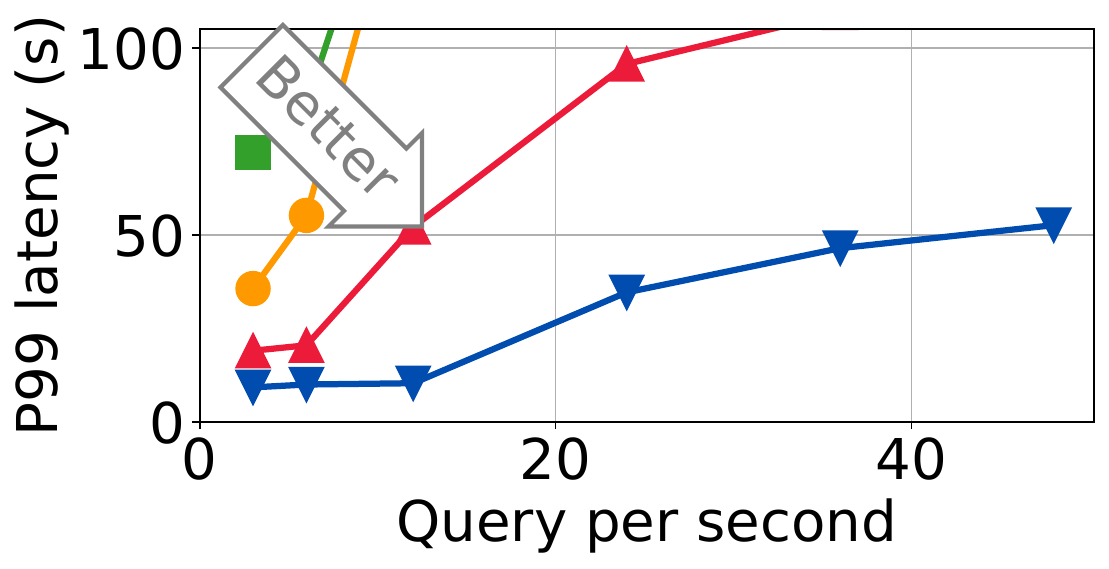}
        \subcaption{Post recommendation\\H100 w/o NVLink\\ QPS --- p99 latency}
    \end{minipage}
    \begin{minipage}{0.24\textwidth}
        \centering
        \includegraphics[width=\linewidth]{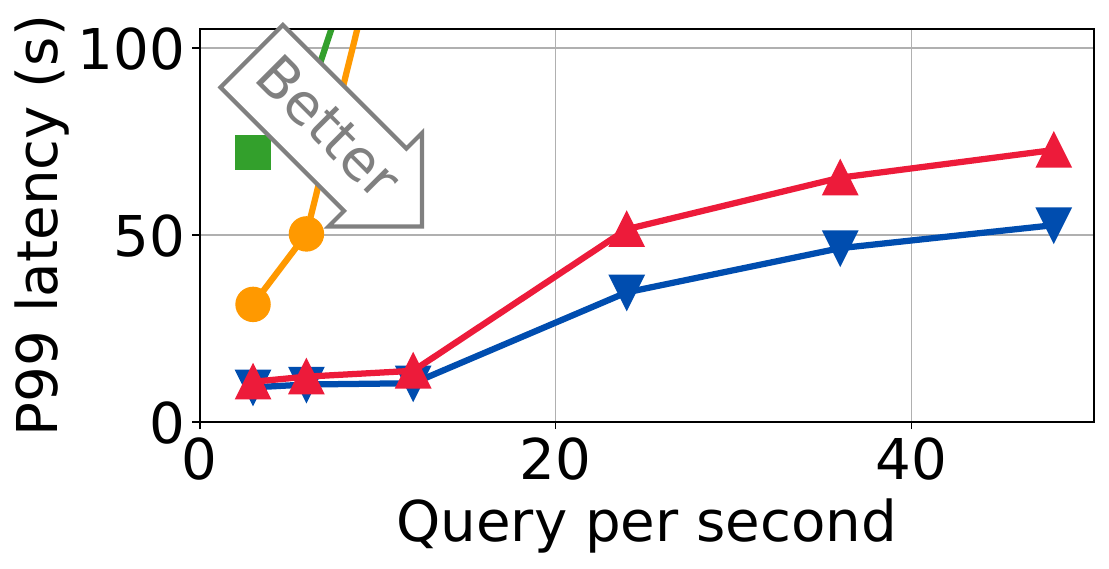}
        \subcaption{Post recommendation\\H100 w/ NVLink\\ QPS --- p99 latency}
    \end{minipage}

    \vspace{0.5em}

    \begin{minipage}{0.24\textwidth}
        \centering
        \includegraphics[width=\linewidth]{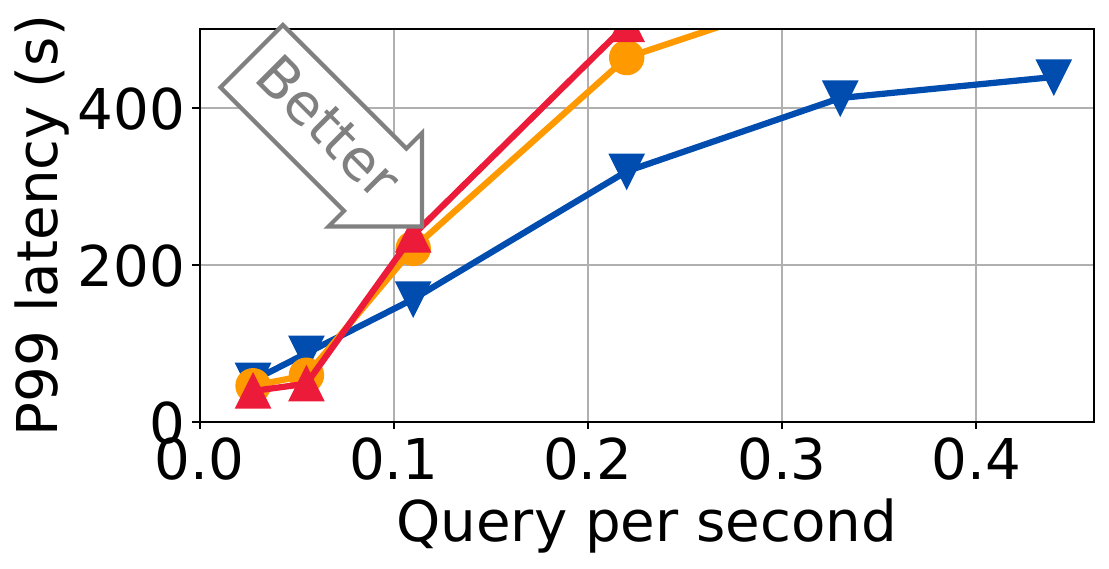}
        \subcaption{Credit verification\\L4\\ QPS --- p99 latency}
    \end{minipage}
    \begin{minipage}{0.24\textwidth}
        \centering
        \includegraphics[width=\linewidth]{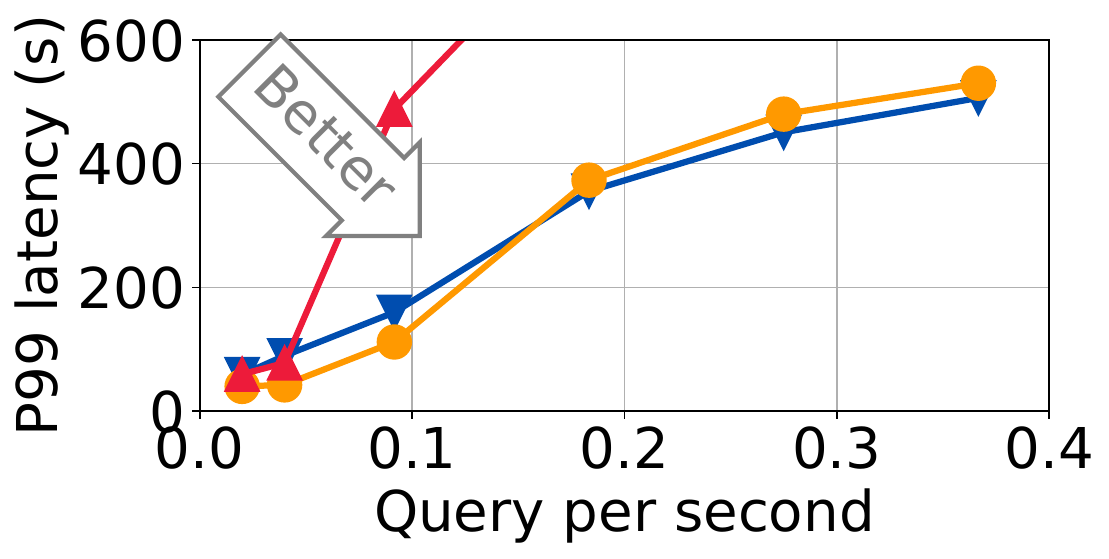}
        \subcaption{Credit verification\\A100\\ QPS --- p99 latency}
    \end{minipage}
    \begin{minipage}{0.24\textwidth}
        \centering
        \includegraphics[width=\linewidth]{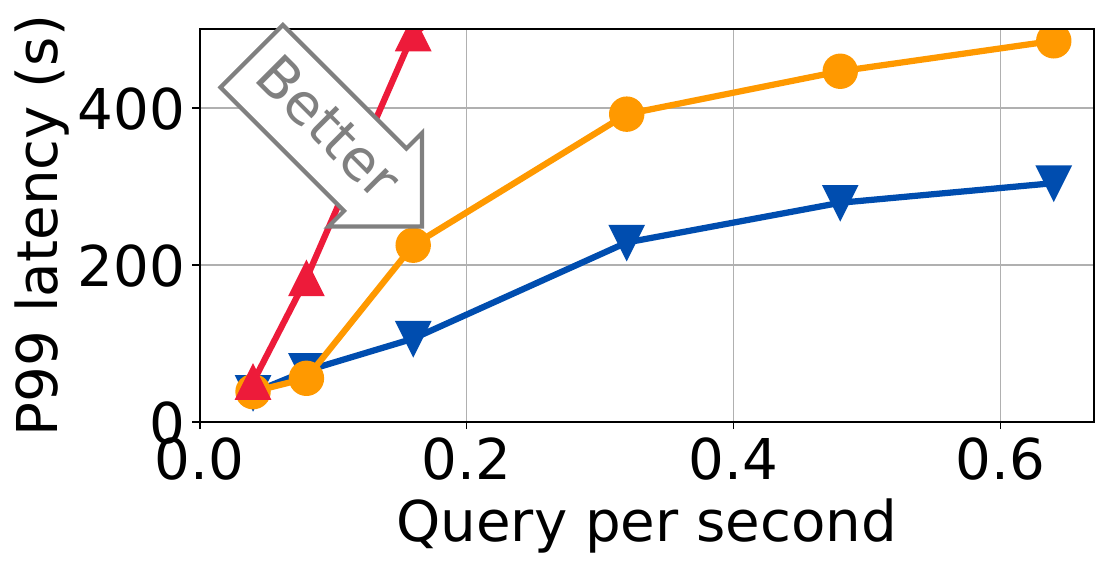}
        \subcaption{Credit verification\\H100 w/o NVLink\\ QPS --- p99 latency}
    \end{minipage}
    \begin{minipage}{0.24\textwidth}
        \centering
        \includegraphics[width=\linewidth]{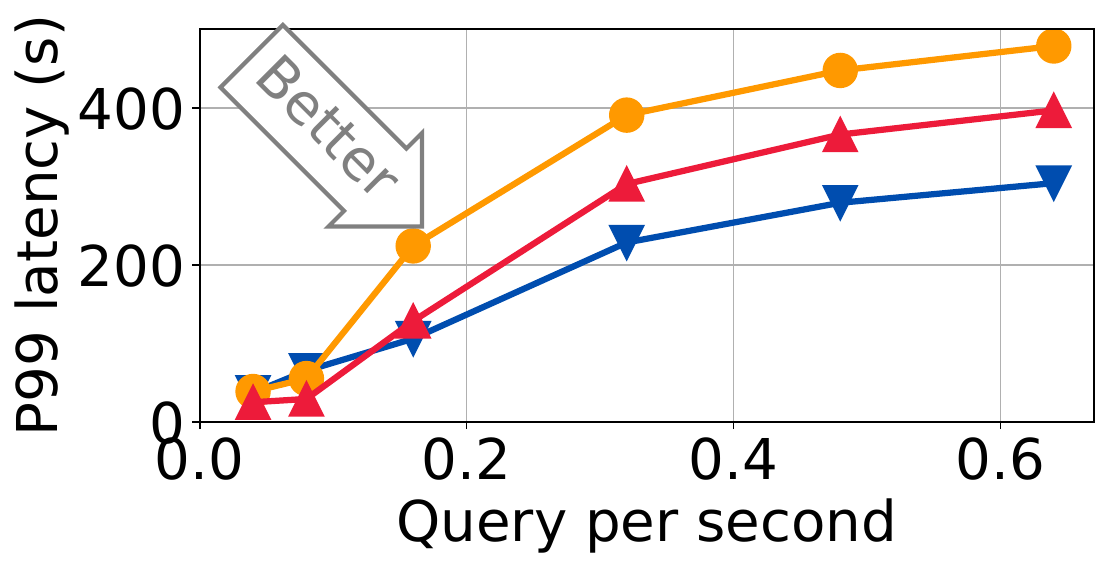}
        \subcaption{Credit verification\\H100 w/ NVLink\\ QPS --- p99 latency}
    \end{minipage}

    \tightcaption{QPS --- P99 latency trade-off of \name and baselines on three different hardware setups and two applications.}
    \label{fig:overall-evaluation-p99-latency}
\end{figure*}

\mypara{Credit verification dataset}
In this dataset, our goal is to simulate a credit verification scenario, where we verify the credit of one user based on the credit history of one user.
We measure that the length of credit history for one month is about 4,000 to 6,000 tokens.
We simulate ten months of credit history, resulting in a credit history length from 40,000 to 60,000 for each user.
We consider 60 users in total.

\mypara{Request arrival pattern}
We assume that the user arrival pattern is a Poisson process. 
We further vary the rate in the Poisson process to vary the query-per-second.

\mypara{Hardware and LLM setup}
We summarize the hardware and the LLM setup in Table~\ref{table:hardware-setup}.

\mypara{\name}
We implement \name based on state-of-the-art LLM serving engine vLLM~\cite{vllm-code-repo}, with 4.6k lines of Python code to implement the core techniques of \name.
We set the fairness parameter $\lambda=500$ by default.

\begin{table}[]
\footnotesize
\begin{tabular}{lll}
\hline
\hline
Scenario                                                 & GPU Type                                                              & LLM Model                                                                                       \\ \hline
\hline
\begin{tabular}[c]{@{}l@{}}Low-end\\ GPU\end{tabular}    & \begin{tabular}[c]{@{}l@{}}2x NVIDIA L4 PCIe\\ (24 GB)\end{tabular}   & \begin{tabular}[c]{@{}l@{}}meta-llama/\\ Llama-3.1-8B\end{tabular}                              \\ \hline
\begin{tabular}[c]{@{}l@{}}Middle-end\\ GPU\end{tabular} & \begin{tabular}[c]{@{}l@{}}2x NVIDA A100 PCIe\\ (40 GB)\end{tabular}  & \begin{tabular}[c]{@{}l@{}}RedHatAI/\\ DeepSeek-R1-Distill-\\ Qwen-32B-FP8-dynamic\end{tabular} \\ \hline
\begin{tabular}[c]{@{}l@{}}High-end\\ GPU\end{tabular}   & \begin{tabular}[c]{@{}l@{}}2x NVIDIA H100 PCIe\\ (80 GB)\end{tabular} & \begin{tabular}[c]{@{}l@{}}Infermatic/\\ Llama-3.3-70B-Instruct-\\ FP8-Dynamic\end{tabular}     \\ 
\hline
\begin{tabular}[c]{@{}l@{}}High-end\\GPU w/\\NVLink\end{tabular}   & \begin{tabular}[c]{@{}l@{}}2x NVIDIA H100 NVLink\\ (80 GB)\end{tabular} & \begin{tabular}[c]{@{}l@{}}Infermatic/\\ Llama-3.3-70B-Instruct-\\ FP8-Dynamic\end{tabular}     \\ 
\hline
\hline
\end{tabular}
\vspace{0.3cm}
\tightcaption{The hardware and the corresponding LLM.}
\label{table:hardware-setup}
\end{table}

\mypara{Baselines}
We pick four baselines, where two of them parallelize the LLM inference (tensor parallel and pipeline parallel) and the other two do not parallelize the inference (PagedAttention~\cite{vllm} and chunked prefill~\cite{chunked-prefill}):
\begin{packeditemize}
    \item \textit{Tensor parallel}. In this baseline, we parallelize the inference onto 2 GPUs with the degree of tensor parallelism equal to 2 using the existing implementation available in production-grade inference engine vLLM~\cite{vllm-code-repo}.
    \item \textit{Pipeline parallel}. In this baseline, we parallelize the inference onto 2 GPUs with the degree of pipeline parallelism equal to 2. We also use the implementation in vLLM~\cite{vllm-code-repo}.
    \item \textit{PagedAttention}~\cite{vllm}. 
    This baseline manages the KV caches using a page table to minimize fragmentation and employs first-come-first-serve scheduling.
    \item \textit{Chunked prefill}~\cite{chunked-prefill}. This baseline processes the LLM input chunk-by-chunk to allow handling longer requests. 
\end{packeditemize}
Note that we enable prefix caching for both \name and all these baselines. Also, some baselines cannot handle some workloads as their maximum input length is too short, we show this in Table~\ref{table:maxlen_wl}

\mypara{Routing}
We note that, for \name and non-parallelization-based baselines, in order to utilize multiple GPUs, we launch multiple instances of LLM inference engines, one on each GPU, and then perform \textit{user-id-based routing}, where we route the request from the same user to the same instance, and decide which user should be assigned to which instance in a round-robin manner.

\begin{figure}

    \centering
    \begin{minipage}{0.49\columnwidth}
    \centering
    \includegraphics[width=\linewidth]{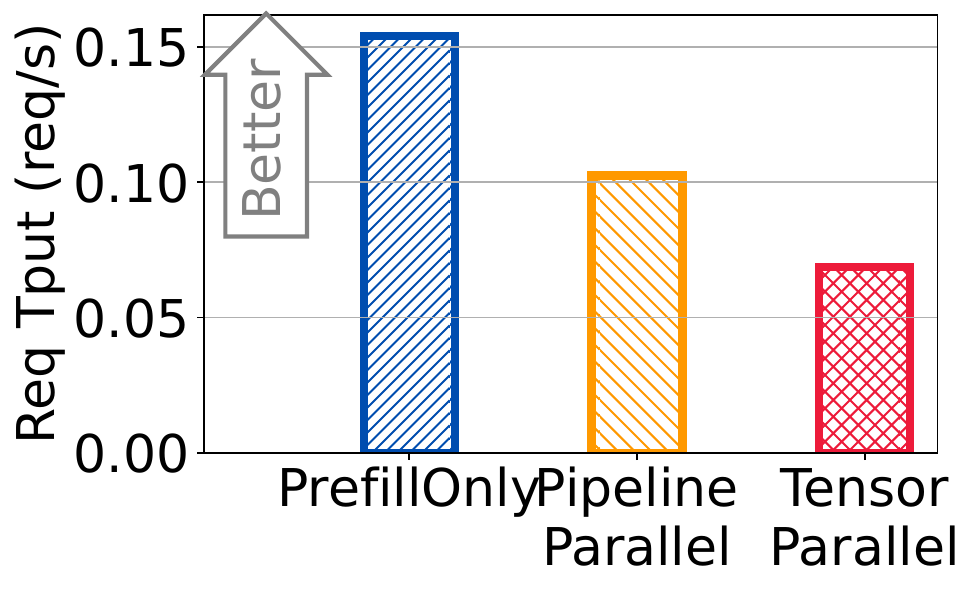}
    \subcaption{Throughput\\w/o NVLink}
    \end{minipage}
    \begin{minipage}{0.49\columnwidth}
    \centering
    \includegraphics[width=\linewidth]{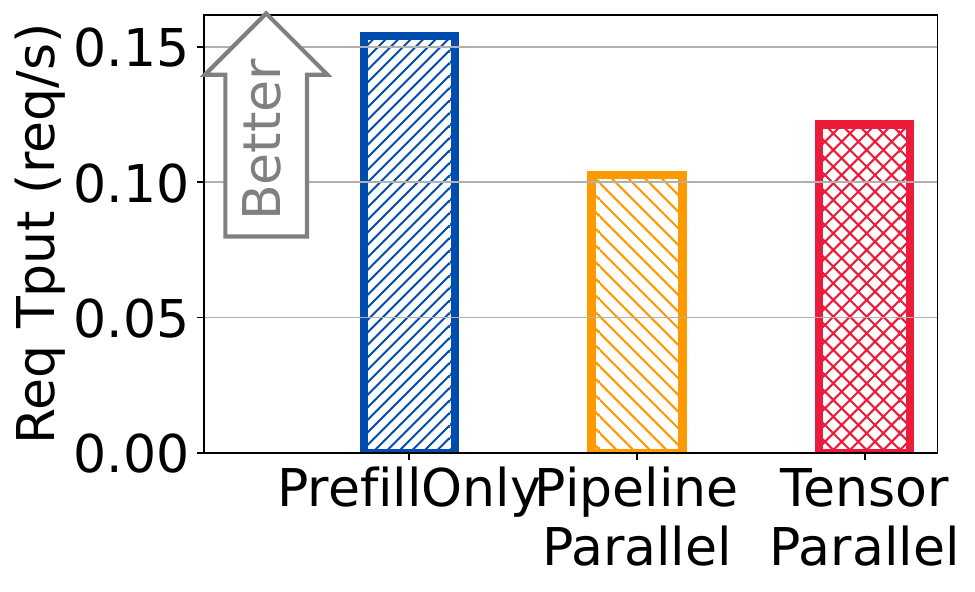}
    \subcaption{Throughput\\w/ NVLink}
    \end{minipage}

    \tightcaption{Contrasting the throughput of \name and baselines on credit verification workload under 2$\times$ H100.
    Though NVLink significantly accelerates the communication and thus enhances the throughput of communication-intensive parallelization like tensor parallel, \name still has the highest throughput as it does not spend extra communication to parallelize the inference.}
    \label{fig:contrast-tput-nvlink}
\end{figure}

\subsection{Evaluation results}

\mypara{QPS--latency trade-off}
We show the trade-off between query per second (QPS) and latency (mean latency and p99 latency) across three different hardware setups and two different applications in Figure~\ref{fig:overall-evaluation-mean-latency}.
In this figure, we determine the evaluation QPS by running \name with all requests in the dataset coming at once, and then obtain the throughput (requests per second) of \name in this situation, where we denote this value as $x$.
We then evaluate QPS $1/4 x, 1/2 x, x, 2x, 3x, 4x$.
This approach allows us to show the full spectrum performance of \name.

In Figure~\ref{fig:overall-evaluation-mean-latency}, we  see that \name always achieves lowest latency when the QPS is high, indicating that the throughput of \name is significantly higher than the baselines.
However, in low QPS the latency of \name can be higher than parallelization-based baselines (as baselines use multiple GPUs to serve one request but \name just uses one).

We also show in Figure~\ref{fig:overall-evaluation-p99-latency} that \name achieves better P99 latency than baselines, indicating that the JCT-based allocation of \name does not hurt P99 latency after applying the fairness twist mentioned in \S\ref{subsec:continuous-jct-calibration}.

\begin{figure}
    \centering
    \includegraphics[width=0.8\columnwidth]{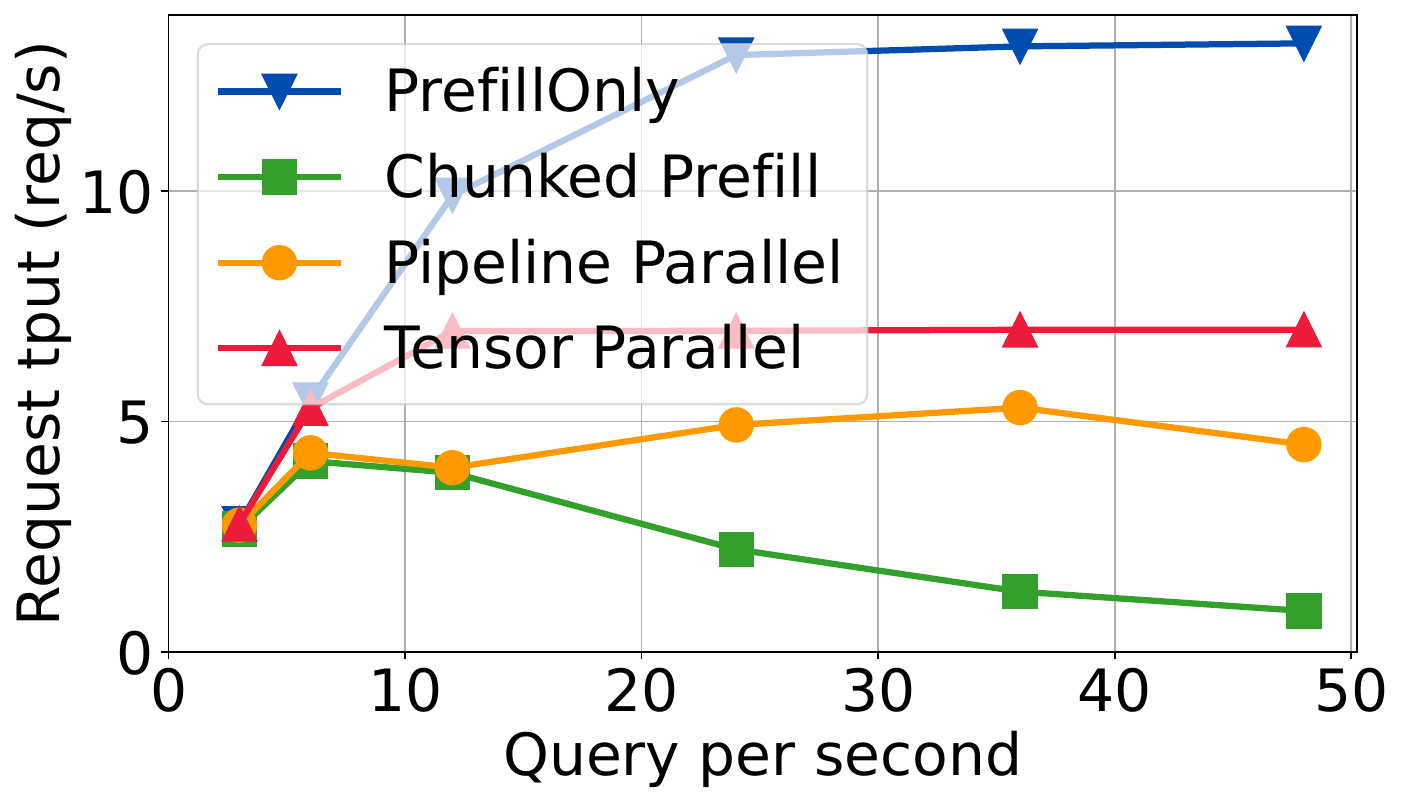}
    \tightcaption{Illustrating the throughput of \name and the baselines in post post-recommendation dataset under 2$\times$ H100 without NVLink. \name has better improvement as it can maintain high throughput under high query-per-second, while the query per second of chunked prefill baseline drops because of prefix cache throttling.
    Parallelization-based baselines parallelize the prefix cache across GPUs and thus have sufficient prefix cache space to avoid throttling, but they have lower throughput because of extra communication and synchronization cost.}
    \label{fig:explain-improvement-tput}
\end{figure}

\mypara{Source of improvement}
We analyze the source of improvement of \name on two datasets.
\begin{packeditemize}
    \item \textit{Post recommendation}
    To understand the source of improvement of \name, we show how the throughput of \name and baselines varies with respect to query per second in Figure~\ref{fig:explain-improvement-tput}.
    We can see that the query per second of chunked prefill baseline drops because the prefix cache throttles under high QPS, and \name avoids such throttling by using continuous JCT calibration to identify and prioritize requests that can hit the prefix cache.
    Parallelization-based baselines parallelize the prefix caches across GPUs and thus have sufficient prefix cache space to avoid throttling, but they have lower throughput because of extra communication and synchronization cost.
    \item \textit{Credit verification}:
    The main reason \name performs better than other baselines under high QPS is because \name can handle long context without parallelizing LLM inference, where a parallelization-based solution will have GPU idle time due to expensive all-reduce communication in the tensor parallel baseline, or the pipeline bubbles in the pipeline parallel baseline.
    To visualize the effect of GPU communication, we contrast the throughput of \name and parallelization solutions between using NVLink and not using NVLink in Figure~\ref{fig:contrast-tput-nvlink}.
    We can see that, though having NVLink can significantly improve the throughput of the tensor parallel baseline due to much faster all-reduce communication, \name still has better throughput as it does not spend extra communication to parallelize the inference.
\end{packeditemize}

\begin{figure}[t]
    \centering
    \includegraphics[width=0.99\linewidth]{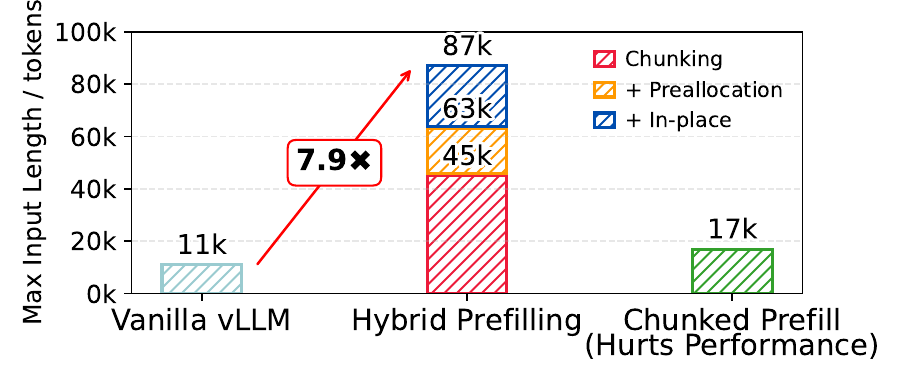}
    \tightcaption{Hybrid prefilling improves the MIL by $7.9\times$ without hurting the throughput, measured on a Qwen-2.5-32B model with fp8 quantization on an A100 GPU.}
    \label{fig:hybrid-prefilling-mil}
\end{figure}

\begin{figure}[t]
    \centering
    \includegraphics[width=0.75\columnwidth]{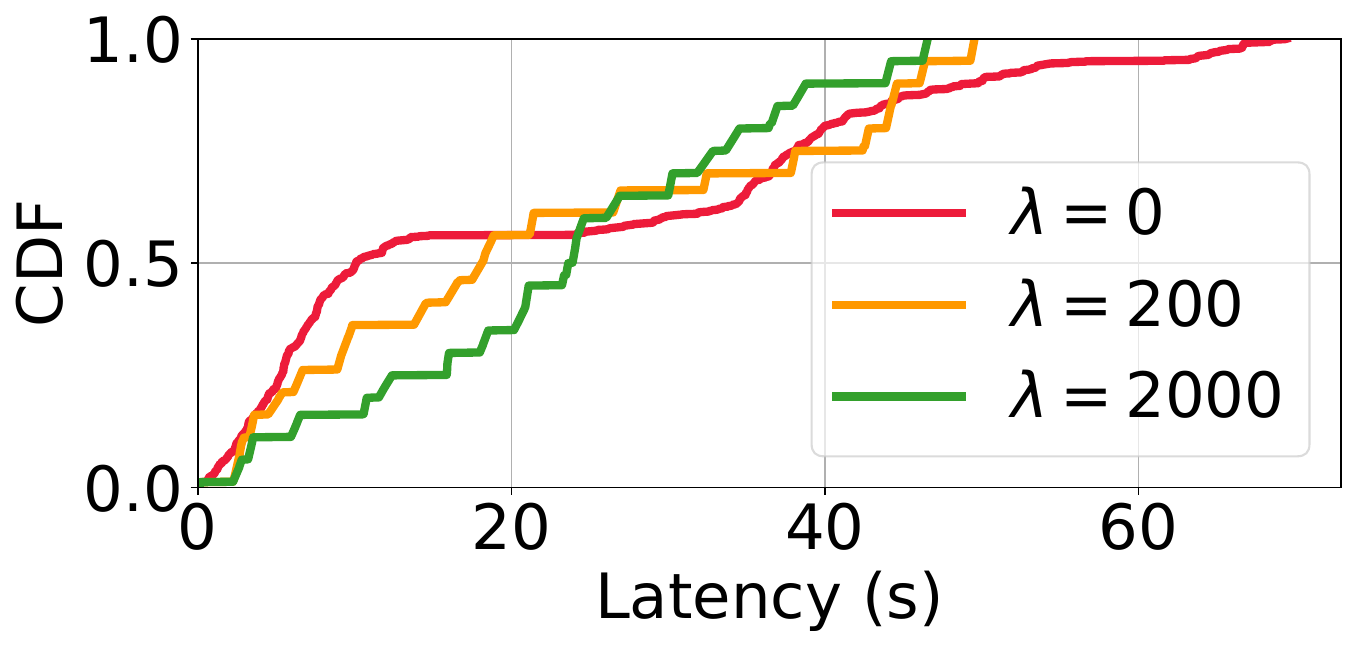}
    \tightcaption{The CDF of request latency of \name, under different value of fairness parameter $\lambda$. Higher $\lambda$ result in better P99 latency, at the cost of inflating the average latency.}
    \label{fig:fairness}
\end{figure}

\mypara{Hybrid prefilling largely improves MIL}
\name can improve the maximum input length of LLM to upto 5$\times$ compared to the non-parallelization-based baseline (as shown in Table~\ref{table:maxlen_wl}) without reducing throughput. Further, to show that hybrid prefilling can effectively enlarge the maximum input length (MIL), we plot how our individual techniques contribute to MIL.
From Figure~\ref{fig:hybrid-prefilling-mil}, we can see that compared to the chunked prefill baseline, \name can improve the maximum context length of a Qwen-2.5-32B model in fp8 precision on an A100 GPU by more than 8.7$\times$.
We note that, though chunked prefill can also improve the MIL, it sacrifices both latency and throughput since it chunks the input and thus reduces GPU kernel efficiency.


\mypara{Varying the fairness parameter $\lambda$}
We vary the CDF of request latency of \name, under different values of fairness parameter $\lambda$ in Figure~\ref{fig:fairness}.
Higher $\lambda$ results in better P99 latency, at the cost of inflating the average latency
\section{Related Work}
\label{sec:related-work}

\mypara{LLM inference engines}
After HuggingFace transformers standardized the way people access open-source LLMs, people started to engineer LLM inference engines that can serve LLMs with high throughput and low latency.
Orca~\cite{orca} introduces continuous batching that schedules LLM inference in token granularity instead of request granularity, allowing the LLM to significantly enlarge the batch size and improve the decoding throughput. 
vLLM~\cite{vllm} further lowers the GPU memory management granularity from request granularity to token granularity, thus minimizing GPU memory fragmentation, which leads to larger batch sizes and thus larger decoding throughput.
DistServe~\cite{distserve} disaggregates prefill and decode across GPUs which significantly improves the goodput under TTFT and TPOT SLO constraints. 

However, these inference engines are suboptimal for prefill-only workload, as the prefill-only request does not require decoding, so fully storing the KV caches unnecessarily enlarges the GPU footprint and does not accelerate this request.
Also, the JCT of prefill-only requests is deterministic, which enables one to improve the scheduling by predicting JCT.


\mypara{LLM caching systems}
Beyond simple prefix caching~\cite{liu2024optimizingsql, jin2024ragcache, zheng2024sglangefficientexecutionstructured}, recent literature started to explore KV cache compression~\cite{cachegen, kvquant, kivi, gear, h2o} to reduce KV cache size, and KV cache blending~\cite{cacheblend, epic, mpic} to reuse non-prefix KV cache.
CacheGen~\cite{cachegen} compresses and streams KV states to reduce loading time for long prompts. CacheBlend~\cite{cacheblend} merges KV caches from multiple documents, selectively recomputing minimal attention states to support efficient RAG-style inference. 
\name is compatible with these works, as \name does not change the KV caches.

\mypara{Building applications via LLM}
Recent studies have started to leverage LLM to build a wide range of applications, including agentic systems~\cite{caravan,generative_agents,autogpt,AgentConf}, recommendation systems~\cite{firooz2025360brew,wang2023enhancing,wu2024survey} and more.
This paper focuses on LLMs for discriminative applications, which typically requires the LLM to generate single output token for long input requests.

\mypara{Traditional deep learning systems}
Prefill-only workload resembles characteristics that are similar to traditional deep learning systems, where traditional deep learning systems also eliminate unnecessary tensors generated during inference~\cite{torch-compile} and leverage JCT-aware scheduling to improve performance~\cite{tiresias,optimus}.
However, the GPU memory footprint of traditional systems may come from non-linear layers like convolution layers, and the JCT is roughly a constant~\cite{reducto,accmpeg,oneadapt,glimpse,eaar}. \name instead embraces LLM-specific properties that the GPU memory footprint mainly comes from linear layers, and the JCT is prefix-caching-sensitive, and develops optimizations on top of these properties.

\section{Discussion}

\mypara{Offloading the KV caches to CPU}
Current implementation of \name performs suffix KV caches discarding, which prevents future requests to potentially reuse the computation of the discarded part.
This limitation can be alleviated by offloading the KV caches to CPU instead via solutions like LMCache~\cite{lmcache}.
We leave this extension to future work.

\mypara{Prefill-decode disaggregation}
Besides prefill-only workload, we found that \name can be used on the prefill node in prefill decode disaggregation scenario, as the workload on the prefill node is also prefill-only.
We leave this extension to future work.

\mypara{Latency-centric optimizations}
\name focuses on exploiting the optimization opportunities in prefill-only workload that improves throughput.
However, we also find that latency can be further optimized for prefill-only requests.
For example, instead of paging the GPU memory~\cite{PagedAttention}, in prefill-only workload we can directly allocate continuous GPU buffers to accelerate GPU computation.

\section{Conclusion}

Besides generative LLM workloads, we observe that LLMs are increasingly being used in traditional discriminative tasks such as recommendation, credit verification, and data labeling.
A new workload --- the prefill-only workload --- emerges under this trend, where LLMs generate only a single output token per request.
Existing LLM engines perform suboptimally on such workloads, as their designs are specialized for multi-token generation.
In this paper, we propose \name, the first LLM inference engine tailored specifically for prefill-only workloads.
\name improves throughput for long requests by enabling one to offload or discard the KV caches during inference without slowing down the inference, thus avoiding parallelizing the KV caches when handling long requests, which hurts throughput.
Furthermore, \name introduces a scheduling algorithm based on continuous calibration of job completion time, reducing latency and improving cache hit rate.
We evaluate \name on 4 different hardware setups, 3 different models, and 2 types of workloads and show that \name can handle up to 4$\times$ higher queries per second without increasing latency, paving the way for empowering day-to-day user applications through recommendation-based LLM pipelines.

\bibliographystyle{plain}
\bibliography{reference}


\end{document}